\documentclass[a4paper,11pt]{article}

\usepackage{jheppub} 

\usepackage[T1]{fontenc} 
\usepackage[utf8]{inputenc}

\usepackage{mathtools}
\usepackage{booktabs,adjustbox}
\usepackage{multirow,dcolumn,makecell}
\usepackage{braket,relsize,slashed}
\usepackage{xcolor}

\let\OLDthebibliography\thebibliography
\renewcommand\thebibliography[1]{
  \OLDthebibliography{#1}
  \setlength{\parskip}{0pt}
  \setlength{\itemsep}{0pt plus 0.3ex}
}

\newcommand{\overbar}[1]{\mkern 1.5mu\overline{\mkern-1.5mu#1\mkern-1.5mu}\mkern 1.5mu}
\newcommand{\pmatr}[1]{\begin{pmatrix} #1 \end{pmatrix}}
\newcommand{\simlt}{~\mbox{\smaller\(\lesssim\)}~}
\newcommand{\simgt}{~\mbox{\smaller\(\gtrsim\)}~}
\newcommand{\Exp}[1]{\exp \left[ #1 \right]}

\title{Flavourful axion phenomenology}
\author[a]{Fredrik~Bj\"{o}rkeroth,}
\author[b]{Eung Jin Chun,}
\author[c]{Stephen F. King}

\affiliation[a]{INFN Laboratori Nazionali di Frascati, Via E. Fermi 40, 00044 Frascati, Italy}
\affiliation[b]{Korea Institute for Advanced Study, Seoul 02455, Korea}
\affiliation[c]{School of Physics and Astronomy, University of Southampton, SO17 1BJ Southampton, United Kingdom}

\emailAdd{fredrik.bjorkeroth@lnf.infn.it}
\emailAdd{ejchun@kias.re.kr}
\emailAdd{king@soton.ac.uk}

\abstract{%
We present a comprehensive discussion of the phenomenology of flavourful axions, including both standard Peccei-Quinn (PQ) axions, associated with the solution to the strong $CP$ problem, and non-standard axion-like particles (ALPs).
We give the flavourful axion-fermion and axion-photon couplings and calculate the branching ratios of heavy meson ($K$, $D$, $B$) decays involving a flavourful axion. 
We also calculate the mixing between axions and heavy mesons $ K^0 $, $ D^0 $, $ B^0 $ and $ B_s^0 $, which affects the meson oscillation probability and mass difference. 
Mixing also contributes to meson decays into axions and axion decays into two photons, and may be relevant for ALPs. 
We discuss charged lepton flavour-violating decays involving final state axions of the form $\ell_1 \to \ell_2 a (\gamma) $, as well as $ \mu \to eee $ and $ \mu-e $ conversion.
Finally we describe the phenomenology of a particular ``A to Z'' Pati-Salam model, in which PQ symmetry arises accidentally due to discrete flavour symmetry.
Here all axion couplings are fixed by a fit to flavour data, leading to sharp predictions and correlations between flavour-dependent observables.
}


\begin{document}
\maketitle
\flushbottom

\section{Introduction}

One of the puzzles of the Standard Model (SM) is why QCD does not appear to break $CP$ symmetry.
The most popular resolution of this so-called ``strong $CP$ problem''
is to postulate a Peccei-Quinn (PQ) symmetry,
namely a QCD-anomalous global $U(1)$ symmetry which is broken spontaneously, leading to a pseudo-Nambu-Goldstone boson (pNGB) called the QCD axion 
\cite{Peccei:1977hh,Wilczek:1977pj,Weinberg:1977ma}.
The two most common approaches to realising such a PQ symmetry is either to introduce heavy vector-like quarks (the KSVZ model) 
\cite{Kim:1979if,Shifman:1979if}
or to extend the Higgs sector (the DFSZ model) 
\cite{Dine:1981rt,Zhitnitsky:1980tq}.
The resulting QCD axion provides a candidate for dark matter 
\cite{Preskill:1982cy,Abbott:1982af,Dine:1982ah}
within the allowed window of the axion (or PQ symmetry-breaking) scale $f_a=10^{9-12}$ GeV 
\cite{Kim:2008hd}.

It has also been realised that the PQ axion need not emerge from an exact global $U(1)$ symmetry, but could result from some discrete symmetry or continuous gauge symmetry leading to an accidental global $U(1)$ symmetry.
Considering the observed accuracy of strong-$CP$ invariance, it is enough to protect the PQ symmetry up to some higher-dimensional operators 
\cite{Holman:1992us,Kamionkowski:1992mf,Barr:1992qq}.
In this regard, it is appealing to consider an approximate PQ symmetry guaranteed by discrete (gauge) symmetries 
\cite{Chun:1992bn,BasteroGil:1997vn,Babu:2002ic,Dias:2002hz,Dias:2002gg,Dias:2007vx,Harigaya:2013vja}.
Alternatively, attempts to link PQ symmetry protected by continuous gauge symmetries to the flavour problem were made in 
\cite{Cheung:2010hk,DiLuzio:2017tjx}.
It is possible that PQ symmetry arises from flavour symmetries 
\cite{Wilczek:1982rv},
linking the axion scale to the flavour symmetry-breaking scale, and various attempts have been made to incorporate such a flavourful PQ symmetry as a part of such \emph{continuous} flavour symmetries 
\cite{Babu:1992cu,Albrecht:2010xh,Celis:2014iua,Ahn:2014gva,Ema:2016ops,Calibbi:2016hwq,Choi:2017gpf,Arias-Aragon:2017eww,Linster:2018avp}.
It is also possible that PQ symmetry could arise accidentally from \emph{discrete flavour symmetries} 
\cite{King:2013eh,King:2015aea,King:2017guk,King:2014nza},
as recently discussed 
\cite{Bjorkeroth:2017tsz} 
in the ``A to Z'' Pati-Salam model 
\cite{King:2014iia},
where quarks and lepton are unified.
This is difficult to achieve in a grand unified theory (GUT) based on $SO(10)$ 
\cite{Bjorkeroth:2015uou}, 
which otherwise presents a stronger case for unification.%
\footnote{
	These ideas should not be confused with alternatives to PQ symmetry, such as Nelson-Barr type resolutions to the strong $ CP $ problem 
	\cite{Nelson:1983zb,Nelson:1984hg,Barr:1984qx,Barr:1984fh},
	or GUT models where specific Yukawa structures have been proposed 
	\cite{Bjorkeroth:2015ora,Antusch:2013rla,Antusch:2013kna}.
}
Recent efforts have been made \cite{Ema:2016ops,Calibbi:2016hwq,Ahn:2018cau} to unify the $ U(1)_{PQ} $ symmetry with a Froggatt-Nielsen-like $ U(1) $ flavour symmetry \cite{Froggatt:1978nt}.
The resultant axion is variously dubbed a ``flaxion'' or ``axiflavon''; we shall refer simply to a ``flavourful axion''.

In this paper we focus on the phenomenology of flavourful axions, including both standard PQ axions, associated with the solution to the strong $CP$ problem, and non-standard axion-like particles (ALPs) (see e.g. \cite{Jaeckel:2013uva}).
For a complementary analysis of ALP signatures and bounds at the LHC, see \cite{Brivio:2017ije}. 
We present the flavourful axion-fermion and axion-photon couplings both for the standard axion and for ALPs, and show that they quite naturally are non-diagonal.
We use these couplings to calculate the branching ratios for two-body decays of heavy mesons $ K $, $ D $, and $ B $ involving a flavourful axion. 
Moreover, we calculate the mixing between axions and neutral hadronic mesons $K^0$, $D^0$, $ B^0 $ and $B_s^0$ and its consequences, which has not been discussed in the literature before.
These can lead to new contributions to neutral meson mass splitting, meson decays into axions and axion decays into two photons which may be relevant for ALPs. 
We also discuss lepton decays involving final state axions, including two-body decays $ \ell_1 \to \ell_2 a $ and radiative decays $ \ell_1 \to \ell_2 a \gamma $, as well as $ \mu \to eee $ and $ \mu-e $ conversion.
Finally we describe the phenomenology of the A to Z Pati-Salam model, which predicts a flavourful axion \cite{Bjorkeroth:2017tsz}, and show how unification leads to correlations between different flavour dependent observables, as the down-type quark and charged lepton couplings are very similar.
Notably, as the axion arises from the same flavon fields that dictate fermion Yukawa structures, no additional field content is necessary to solve the strong $ CP $, and all axion couplings are fixed by a fit to quark and lepton masses and mixing.

The layout of the remainder of the paper is as follows.
Section~\ref{sec:couplings} describes the flavourful axion-fermion and axion-photon couplings both for the standard axion and for ALPs.
In Section~\ref{sec:mesondecays} we apply these couplings to calculate the branching ratios of heavy meson decays involving a flavourful axion. 
Section~\ref{sec:mesonmixing} discusses the mixing between axions and neutral mesons while
Section~\ref{sec:leptondecays} discusses lepton decays.
Section~\ref{sec:AtoZ} focusses on the phenomenology of the A to Z model, which predicts correlations between different flavour dependent observables, and Section~\ref{sec:conclusion} concludes.
Appendix~\ref{app:sec:mesonmixing} gives more details about axion-meson mixing.
Appendix~\ref{app:sec:br} details the calculation the heavy meson branching ratios.
Appendix~\ref{app:sec:couplings} shows the derivation of the couplings in the A to Z Pati-Salam model and 
Appendix~\ref{app:sec:numericalfit} tabulates the numerical fit to flavour data.

\section{Axion couplings to matter}
\label{sec:couplings}

\subsection{Lagrangian}

Relevant to a discussion on axion-fermion interactions is the Lagrangian 
\begin{equation}
	\mathcal{L} = 
		\mathcal{L}_\mathrm{kin} 
		+ \mathcal{L}_m 
		+ \mathcal{L}_\partial 
		+ \mathcal{L}_\mathrm{anomaly} ,
\end{equation}
where 
$ \mathcal{L}_\mathrm{kin} $ contains the kinetic terms,
$ \mathcal{L}_m $ the fermion mass terms,
$ \mathcal{L}_\partial $ the axion derivative couplings to matter, and
$ \mathcal{L}_\mathrm{anomaly} $ the QCD and electromagnetic anomalies.
In the physical (mass) basis below the electroweak symmetry-breaking scale, we have
\begin{equation}
\begin{split}
	\mathcal{L}_\mathrm{kin} + \mathcal{L}_m &=
		\frac{1}{2} (\partial_\mu a)^2
    	+ \sum_{f = u, d, e} 
		\bar{f}_i (\slashed{\partial} - m_i) f_i 
		, \\
	\mathcal{L}_\partial &=  
		- \frac{\partial_\mu a}{v_{PQ}} \sum_{f = u, d, e} 
		\bar{f}_i \gamma^\mu (V^f_{ij} - A^f_{ij} \gamma_5) f_j , \\
	\mathcal{L}_\mathrm{anomaly} &=
		\frac{\alpha_s}{8\pi} \frac{a}{f_a} G_{\mu\nu}^a \tilde G^{a\mu\nu}
		+ c_{a\gamma} \frac{\alpha}{8\pi} 
		\frac{a}{f_a} F_{\mu\nu} \tilde F^{\mu\nu} ,
\end{split}
\label{eq:lagrangian}
\end{equation}
with the axion decay constant $ f_a = v_{PQ} / N_{DW} $ defined in terms of the PQ-breaking scale $ v_{PQ} $ and anomaly (or domain wall) number $ N_{DW} $.
The axion-photon coupling is discussed in Section~\ref{sec:axionparameters} below.
The physical masses $ m^f_i $ are defined by $ m^f_i = (U_{Lf}^\dagger M^f U_{Rf})_{ii} $, in terms of the mass matrix in the weak basis, $ M^f $, and unitary matrices $ U_{Lf} $, $ U_{Rf} $ which transform left- and right-handed fields, respectively.
The vector and axial couplings are given by
\begin{equation}
\begin{split}
	V^f &= \frac{1}{2} (X_L + X_R) = \frac{1}{2} \left( U_{Lf}^\dagger x_{f_L} U_{Lf} + U_{Rf}^\dagger x_{f_R} U_{Rf} \right) , \\
	A^f &= \frac{1}{2} (X_L - X_R) = \frac{1}{2} \left( U_{Lf}^\dagger x_{f_L} U_{Lf} - U_{Rf}^\dagger x_{f_R} U_{Rf} \right) .
\end{split}
\label{eq:VfAf}
\end{equation}
$ x_{f_L} $, $ x_{f_R} $ are the fermion PQ charges in the left-right (LR) basis,%
\footnote{
	Note that right-handed particles in supersymmetric theories typically manifest as left-handed antifermions $ f^c $. 
	Then $ x_{f^c} \equiv - x_{f_R} $, where $ x_{f^c} $ is the PQ charge in the ``SUSY basis'' where the superpotential is defined.
} 
written here as (diagonal) matrices.
As $ x_{f_L} $, $ x_{f_R} $ are real, $ V^f $ and $ A^f $ (as well as chiral coupling matrices $ X_{L,R} $) are Hermitian.

In this formulation, the implications of flavour structure are clear.
If all generations of a fermion couple equally to the axion, the charge matrices $ x_{Lf,Rf} $ are proportional to the identity, i.e. 
$ V^f = \frac{1}{2} (x_{f_L} + x_{f_R}) \mathbb{I}_3 $, 
$ A^f = \frac{1}{2} (x_{f_L} - x_{f_R}) \mathbb{I}_3 $,
and there is no flavour violation.
In standard axion models, e.g. DFSZ, charges can be assigned such that $ x_{f_L} = - x_{f_R} $ and the axion couples only via $ A^f $; this is generally not true in flavoured axion models.
Meanwhile if $ x_{f_L} = x_{f_R} $, the $ U(1)_{PQ} $ transformation is not chiral ($ N_{DW} = 0 $), the Goldstone field $ a $ doesn't couple to the QCD anomaly, the strong $ CP $ problem is not solved, and $ a $ is then interpreted as an ALP.%
\footnote{
	The mass of the ALP no longer arises from the QCD vacuum, and the relation $ m_a \propto 1/f_a $ no longer holds.
	We don't specify any particular mass generation scheme here.
}
However, as long as $ x_{f_L,f_R} \slashed{\propto} \mathbb{I}_3 $, we still get flavour-violating (vector and axial) interactions due to weak mixing encoded in $ U_{Lf,Rf} $.

\subsection{Physical axion basis}
\label{sec:physicalaxionbasis}

The above Lagrangian describes an interacting axion, not necessarily in its mass eigenstate.
The off-diagonal couplings to fermions are nevertheless $ V^f $ and $ A^f $ for the physical axion, as we will see.
Unlike standard DFSZ models with PQ-charged Higgs doublets, our flavoured axion does not mix with the longitudinal component of the $ Z $ boson.
We still need to identify the physical axion at low energy as the state orthogonal to $ \pi^0 $ and $ \eta $ mesons.
One can then determine the canonical axion mass and couplings 
\cite{Bardeen:1977bd,Srednicki:1985xd,Bardeen:1986yb}. 
Let us briefly summarize how it works, following the prescription e.g. in 
\cite{Kim:2008hd}. 
The axion mass generated by the QCD anomaly coupling in Eq.~\ref{eq:lagrangian} is conveniently calculated by rotating away the anomaly via chiral transformations of light quarks ($q=u,d,s$),
\begin{equation}
	q \to e^{i \frac{\beta_q}{2} \frac{a}{f_a} \gamma_5} q , \qquad 
	\beta_q = \frac{m_\ast}{m_q} ,
\label{eq:axialtransformation}
\end{equation} 
where $ m_\ast^{-1} = m_u^{-1} + m_d^{-1} + m_s^{-1}$. 
For $ m_{u,d} \ll m_s $ (a good approximation to leading order), we have $ m_*^{-1} \approx m_u^{-1}+m_d^{-1} $.
This leads to a low-energy effective Lagrangian below the chiral symmetry-breaking scale, 
\begin{equation}
	\mathcal{L}_\mathrm{eff} \supset 
		-m_u \braket{\bar{u}_L u_R} e^{i \left( \frac{\pi^0}{f_\pi} + \beta_u \frac{a}{f_a} \right)}
		-m_d \braket{\bar{d}_L d_R} e^{i \left(-\frac{\pi^0}{f_\pi} + \beta_d \frac{a}{f_a} \right)} 
		+ \mathrm{h.c.} .
\label{eq:Leff}
\end{equation}
Using the relation $ \braket{\bar{u}_L u_R} = \braket{\bar{d}_L d_R} = m_\pi^2 f_\pi^2/(m_u+m_d)$, the axion-pion mixing term vanishes.
We identify the state $ a $ in Eq.~\ref{eq:Leff} as the physical axion and extract its mass,
\begin{equation}
	m_a^2 = \frac{m_u m_d}{(m_u + m_d)^2} \frac{m_\pi^2 f_\pi^2}{f_a^2} .
\end{equation}
There remains additional mixing with heavier mesons such as $ \eta^\prime $ which provide further small corrections.
A precise calculation performed in \cite{diCortona:2015ldu} gives us
\begin{equation}
	m_a = 5.70(6)(4) \left( \frac{10^{12} \mathrm{~GeV}}{f_a} \right) \mathrm{~\mu eV}.
\label{eq:ma}
\end{equation}

The transformation in Eq.~\ref{eq:axialtransformation} affects also the axion-quark couplings.
For example for $ u $, $ d $ and $ s $ quarks, the axion-quark Lagrangian in Eq.~\ref{eq:lagrangian} is transformed into the physical basis,
\begin{equation}
	\mathcal{L}_\partial \to
	\mathcal{L}'_\partial \supset
		- \frac{\partial_\mu a}{v_{PQ}} 
		\left[
			\sum_{q=u,d,s} c_q \bar{q} \gamma^\mu \gamma_5 q 
      		+ \bar{s} \gamma^\mu (c^{\prime}_{sd} - c_{sd} \gamma_5) d 
    		+ \bar{d} \gamma^\mu (c^{\prime \ast}_{sd} - c_{sd}^\ast \gamma_5) s
     	\right] ,
\label{eq:aqq2}
\end{equation}
where
$ c_u = A^u_{11} + N_{DW} \beta_u/2$, 
$ c_d = A^d_{11} + N_{DW} \beta_d/2$, 
$ c_s = A^d_{22} + N_{DW} \beta_s/2$,
$ c_{sd}^\prime = V^d_{21}$,
and 
$ c_{sd} = A^d_{21}$. 
We see that the diagonal couplings are modified by an amount proportional to $ N_{DW} $, whereas the off-diagonal couplings are unchanged.
Physically, this is a consequence of the QCD anomaly being flavour-conserving, and unable to mediate flavour-violating interactions that contribute to $ c_{sd} $.

The above discussion identifies the physical axion basis in the limit of no kinetic mixing between the axion and heavier mesons. 
Such mixing, induced by the effective Lagrangian in Eq.~\ref{eq:aqq2}, needs to be further diagonalized away to obtain the physical axion basis.
This will be discussed in detail in Section~\ref{sec:mesonmixing} and Appendix~\ref{app:sec:mesonmixing}.
The kinetic mixing contribution is negligibly small for the standard QCD axion with $m_a \lll m_\pi$ and $f_a \ggg f_\pi$, but can be important for an ALP.

\subsection{Decay constant and axion-photon coupling}
\label{sec:axionparameters}

In standard axion scenarios, the decay constant $ f_a $ is defined by $ v_{PQ}/N_{DW} $, where $ N_{DW} $ is the QCD anomaly number. 
Provided the $ U(1)_{PQ} $ symmetry is broken by the VEV of a single field $ \phi $ with PQ charge $ x_\phi $, we simply have $ v_{PQ} = x_\phi v_\phi $.%
\footnote{
	Its PQ charge $ x_\phi $ can be removed by normalising all charges such that $ x_\phi \equiv 1 $.
}
In more general models, where several fields $ \phi $ contribute to symmetry breaking, we define $ v_{PQ}^2 = \sum_\phi x_\phi^2 v_\phi^2 $. 
If one VEV $ v_{\phi_i} $ dominates, we recover to good approximation the one-field relation; if, say, $ v_{\phi_{j \neq i}} \simlt 0.1 v_{\phi_i} $, $ v_{PQ} \approx x_{\phi_i} v_{\phi_i} $ to within 1\%.
We will encounter exactly this scenario when discussing the A to Z model presented in Section~\ref{sec:AtoZ}.

The axion-photon coupling $ a F \tilde{F} $ defined in Eq.~\ref{eq:lagrangian} is given in terms of the electromagnetic anomaly number $ E $, through the coefficient
\begin{equation}
	c_{a\gamma} = \frac{E}{N_{DW}} - \frac{2(4+z)}{3(1+z)} , \qquad
	z = \frac{m_u}{m_d} \approx 0.56 .
\label{eq:axionphotoncoupling}
\end{equation}
In unified models, such as the A to Z model with Pati-Salam unification presented in Section~\ref{sec:AtoZ}, the ratio of anomaly numbers is fixed to $ E/N_{DW} = 8/3 $, giving $ c_{a\gamma} \approx 0.75 $.

\section{Heavy meson decays}
\label{sec:mesondecays}

The flavour-changing vector couplings in $ \mathcal{L}_\partial $ may lead to observable decays of heavy mesons into axions.
A general study of such flavour-changing processes involving a (massless) Nambu-Goldstone boson was made in \cite{Feng:1997tn}, which is applicable to our flavourful axion.

For a two-body decay $ P \to {P^\prime} a $ of a heavy meson $ P = (\bar{q}_P q^\prime) $ into $ {P^\prime} = (\bar{q}_{P^\prime} q^\prime) $, the branching ratio is given by
\begin{equation}
	\mathrm{Br}(P \to {P^\prime} a) = 
		\frac{1}{16 \pi \Gamma(P)}
		\frac{\big| V^f_{q_P q_{P^\prime}} \big|^2}{v_{PQ}^2}
		m_P^3
		\left( 1 - \frac{m_{P^\prime}^2}{m_P^2} \right)^3
		\left| f_{+}(0) \right|^2,
\label{eq:BrPtoYa}
\end{equation}
with $ V^f $ as defined in Eq.~\ref{eq:VfAf}.
Its indices $ {q_P q_{P^\prime}} $ relate to the constituent quarks, e.g. a $ K^+ \to \pi^+ a $ decay proceeds by $ \bar{s} \to \bar{d} a $ with coupling strength $ V^d_{sd} \equiv V^d_{21} $.
For completeness, a rederivation of Eq.~\ref{eq:BrPtoYa} is provided in Appendix~\ref{app:sec:br}.
It depends on a form factor $ f_{+}(q^2) $ encapsulating hadronic physics, where $ q = p_a = p_P - p_{P^\prime} $ is the momentum transfer to the axion. 
The lightness of the axion means we can safely take the limit $ q^2 \to 0 $.
For kaon decays, $ f_{+}(0) \approx 1 $ to good approximation. 
For heavier mesons, we use results from lattice QCD \cite{AlHaydari:2009zr}, summarised in Table~\ref{tab:fplus}.

\begin{table}[!ht]
\centering
\begin{tabular}{D{-}{\ \to \ }{3,3}c}
\toprule
	\multicolumn{1}{c}{Decay} 		& $ f_{+}(0) $ \\
\midrule
	K - \pi 	& 1			 \\
	D - \pi 	& 0.74(6)(4) \\
	D - K 		& 0.78(5)(4) \\
	D_s - K 	& 0.68(4)(3) \\
	B - \pi 	& 0.27(7)(5) \\
	B - K 		& 0.32(6)(6) \\
	B_s - K 	& 0.23(5)(4) \\
\bottomrule
\end{tabular}
\caption{
	Form factors $ f_{+}(0) $ extracted from \cite{AlHaydari:2009zr} for $ K $, $ D $ and $ B $ decays.
}
\label{tab:fplus}
\end{table}

\subsubsection*{\boldmath{$K^+ \to \pi^+ a$}}

The canonical example of this type of flavour-violating decay is $ K^+ \to \pi^+ a $, which can be constrained by searches for the rare decay $ K^+ \to \pi^+ \nu \bar{\nu} $.
This was done in the E949 and E787 experiments, which observed in total seven events.
Combined analyses \cite{Artamonov:2009sz} (see also \cite{Adler:2008zza}) yield a measurement
$ \mathrm{Br}(K^+ \to \pi^+ \nu \bar{\nu}) = 1.73^{+1.15}_{-1.05} \times 10^{-10} $,
consistent with the SM prediction
$ (0.84 \pm 0.10) \times 10^{-10} $ \cite{Buras:2015qea}.
A bound on axion decays is also provided: 
$ \mathrm{Br}(K^+\to \pi^+ a) < 0.73 \times 10^{-10} $ at 90\% CL \cite{Adler:2008zza}.
The current NA62 experiment at CERN, which recently recorded their first $ K^+ \to \pi^+ \nu \bar{\nu} $ event \cite{MarchevskiMoriond2018}, is expected to observe over $ 100 $ events, reaching a sensitivity of 
$ \mathrm{Br}(K^+\to \pi^+ a) < 1.0 \times 10^{-12} $ at 90\% CL \cite{Fantechi:2014hqa}.

\subsubsection*{\boldmath{$K^0_L \to \pi^0 a$}}

Searches have also been performed for the neutral kaon decay $ K^0_L \to \pi^0 \nu \bar{\nu} $, for which the SM predicts $ \mathrm{Br}(K^0_L \to \pi^0 \nu \bar{\nu}) = (2.9 \pm 0.2) \times 10^{-11} $ 
\cite{Gorbahn:2011pd,Brod:2010hi,Antonelli:2009ws}.
The current best limit is set by the E391a experiment at KEK, giving $ \mathrm{Br}(K^0_L \to \pi^0 \nu \bar{\nu}) < 2.6 \times 10^{-8} $ at 90\% CL \cite{Ahn:2009gb}.
Its successor KOTO has been constructed at J-PARC.
A pilot run in 2013 yielded a limit $ \mathrm{Br}(K^0_L \to \pi^0 \nu \bar{\nu}) < 5.1 \times 10^{-8} $ \cite{Beckford:2017gsf}.
An analysis has also been performed for $ K^0_L \to \pi^0 X^0 $ for a boson $ X^0 $ of arbitrary mass.
For $ m_{X^0} \simeq 0 $, they set $ \mathrm{Br}(K^0_L \to \pi^0 X^0) \simlt 5 \times 10^{-8} $ \cite{Ahn:2016kja}.
Detector upgrades and additional data taken since 2015 are expected to significantly improve these bounds.
An additional experiment dubbed KLEVER has been proposed to measure $ K^0_L \to \pi^0 \nu \bar{\nu} $ at the CERN SPS \cite{Moulson:2016zsl}.

\subsubsection*{\boldmath{$B$} and \boldmath{$B_s$} decays}

$ B $ physics has a rich phenomenology, and is recently of particular interest due to persistent anomalies in observed semileptonic $ B $ decays at the LHC, which may be evidence for charged lepton flavour violation (cLFV) \cite{Buttazzo:2017ixm}.
Rare $ B $ decays of the type $ B \to \pi (K) \nu \bar{\nu} $, while generally not as tightly constrained as those for kaons, may also provide insights into new physics.
A dedicated search for decays like $ B \to \pi (K) a $  with a light invisible particle $ a $ was made by CLEO, which collected $ 10^7 $ $ B \overbar{B} $ pairs throughout its lifetime.
It provides the limits
$ \mathrm{Br}(B^\pm \to \pi^\pm (K^\pm) a) < 4.9 \times 10^{-5} $ and 
$ \mathrm{Br}(B^0 \to \pi^0 (K^0) a) < 5.3 \times 10^{-5} $ at 90\% CL
\cite{Ammar:2001gi}.
More recent and powerful experiments, namely BaBar and Belle, have not yet provided limits on this exact process.
However we may estimate their experimental reach by the stated limits on the decays $ B \to \pi (K) \nu \bar{\nu} $, which are typically $ \mathcal{O}(10^{-5}) $ (see Table~\ref{tab:2bodydecays}), an improvement of approximately one order of magnitude.
 The upgraded experiment Belle-II at SuperKEKB is expected to collected approximately $ N = 5 \times 10^{10} $ $ B \overbar{B} $ pairs, improving the limits on many rare decays \cite{Cunliffe:2017cox}; 
assuming the sensitivity scales as $ \sqrt{N} $, we may expect an $ \mathcal{O}(10^2) $ improvement in branching ratio limits.

It is worth noting that the decay $ B^0 \to \pi^0 a $, predicted by flavoured axion models, has not been analysed explicitly by experiments. 
However, some information may be gleaned from searches for the SM process $ B^0 \to \pi^0 \nu \bar{\nu} $, which are a background to the axion signal.
Generically, any bound on the SM decay will translate into a bound as strong (or stronger) on the two-body decay to an axion.
Finally, we remark on the fact that also decays of the form $ B_s^0 \to \overbar{K}^0 a $ and $ B_s^0 \to \eta (\eta^\prime) a $ are allowed, but no meaningful experimental information is available.

\subsubsection*{\boldmath{$D$} and \boldmath{$D_s$} decays}

Little is said in the literature about decays of charmed mesons of the form $ D^{\pm,0} \to \pi^{\pm,0} a $ or $ D_s^\pm \to K^\pm a $, or the corresponding decays involving a $ \nu \bar{\nu} $ pair.
The branching ratio for $ D \to \pi(K) a $ may be easily calculated using the same formulas for $ K $ and $ B $ decays, given below.
The trivial requirement that $ \mathrm{Br}(D \to \pi (K) a) < 1 $ allows us to place weak bounds on $ v_{PQ} $ of $ \mathcal{O}(100) $ TeV, but without an experimental probe, little more can be said. 
As we will show below, the predicted branching ratios are anyway expected to be rather small when compared to $ K $ and $ B $ decays, which have corresponding branching ratios approximately three and one order of magnitude greater. 
In conclusion, while further experimental probes of $ D $ decays are of course welcome, they are not expected to be more sensitive to flavoured axions than other decays.
On the other hand, in flavoured axion scenarios only $ D $ decays can probe the up-type quark Yukawa matrix.

\subsubsection*{Bounds}

Ultimately the experimental data can be used to constrain the ratio $ |V^f_{q_P q_{P^\prime}}| / v_{PQ} $ for a given decay.
Collecting terms in Eq.~\ref{eq:BrPtoYa}, we define a branching ratio coefficient $ \tilde{c}_{P \to {P^\prime}} $, which depends only on hadronic physics, by
\begin{equation}
	\mathrm{Br}(P \to {P^\prime} a) = 
		\tilde{c}_{P \to {P^\prime}} \left| V^f_{q_P q_{P^\prime}} \right|^2 \left(\frac{10^{12} \mathrm{~GeV}}{v_{PQ}}\right)^2 ,
\label{eq:BrPYcPY}
\end{equation}
i.e.
\begin{equation}
	\tilde{c}_{P \to {P^\prime}} =
		\frac{1}{16\pi \, \Gamma(P)}
		\frac{m_P^3}{(10^{12} \mathrm{~GeV})^2}
		\left( 1 - \frac{m_{P^\prime}^2}{m_P^2} \right)^3
		\left| f_{+}(0) \right|^2.
\label{eq:cPY}
\end{equation}
The values of $ \tilde{c}_{P \to {P^\prime}} $ are tabulated in Table~\ref{tab:2bodydecays}, along with experimental limits on the branching ratio and the corresponding bound on $ v_{PQ} $, where available.
$ D $, $ D_s $ and $ B_s $ decays have no experimental constraints, however we can compute the numerical coefficients $ \tilde{c} $, which are all $\mathcal{O}(10^{-14} - 10^{-13})$.
These are also given in Table~\ref{tab:2bodydecays}.

\begin{table}[!ht]
\centering
\begin{tabular}{D{-}{\ \to \ }{3,3} rrrrr}
\toprule
	\multicolumn{1}{c}{Decay} & 
	Branching ratio & 
	Experiment &
	$ \tilde{c}_{P \to {P^\prime}} $ &
	$ v_{PQ} /\mathrm{GeV} $ \\
\midrule
	K^+ - \pi^+ a 
	& $ \mathbf{< 0.73 \times 10^{-10} } $ 	& E949 + E787 \cite{Adler:2008zza} 		& $ 3.51 \times 10^{-11} $ 	& $ > 6.9 \times 10^{11} |V^d_{21}| $ \\
	& $ < 0.01 \times 10^{-10} $* 			& NA62 (future) \cite{Fantechi:2014hqa}	& 							& $ > 5.9 \times 10^{12} |V^d_{21}| $ \\
	& $ < 1.2 \times 10^{-10} $				& E949 + E787 \cite{Artamonov:2009sz}	& 							& \\
	& $ < 0.59 \times 10^{-10} $			& E787 \cite{Adler:2001xv} 				& 							& \\
\midrule[0.2pt]
	K_L^0 - \pi^0 a 
	& $ \mathbf{< 5 \times 10^{-8}}$ 		& KOTO \cite{Ahn:2016kja} 				& $ 3.67 \times 10^{-11} $	& $ > 2.7 \times 10^{10} |V^d_{21}| $\\
	(K_L^0 - \pi^0 \nu \bar{\nu})
	& $ ( < 2.6 \times 10^{-8} ) $   		& E391a \cite{Ahn:2009gb}				&  							&  \\
\midrule
	B^\pm - \pi^\pm a 
	& $ \mathbf{< 4.9 \times 10^{-5}} $		& CLEO \cite{Ammar:2001gi} 				& $ 5.30 \times 10^{-13} $ 	& $ > 1.0 \times 10^8 |V^d_{31}| $ \\
	(B^\pm - \pi^\pm \nu \bar{\nu})
	& $ (< 1.0 \times 10^{-4} ) $			& BaBar \cite{Aubert:2004ws}			& & \\
	& $ (< 1.4 \times 10^{-4} ) $			& Belle \cite{Grygier:2017tzo}			& & \\
\midrule[0.2pt]
	B^\pm - K^\pm a 
	& $ \mathbf{< 4.9 \times 10^{-5}} $ 	& CLEO \cite{Ammar:2001gi} 				& $ 7.26 \times 10^{-13} $ 	& $ > 1.2 \times 10^8 |V^d_{32}| $ \\
	(B^\pm - K^\pm \nu \bar{\nu})
	& $ (< 1.3 \times 10^{-5} ) $	 		& BaBar \cite{delAmoSanchez:2010bk}		& & \\
	& $ (< 1.9 \times 10^{-5} ) $ 			& Belle \cite{Grygier:2017tzo} 			& & \\
	& $ (< 1.5 \times 10^{-6} ) $* 			& Belle-II (future) \cite{Abe:2010gxa}  & & \\
\midrule[0.2pt]
	B^0 - \pi^0 a 
	&										&										& $ 4.92 \times 10^{-13} $ 	& \\
	(B^0 - \pi^0 \nu \bar{\nu}) 
	& $ ( < 0.9 \times 10^{-5} ) $ 			& Belle \cite{Grygier:2017tzo}			&							& $ \simgt 2.3 \times 10^8 |V^d_{31}| $ \\
\midrule[0.2pt]
	B^0 - K^0_{(S)} a 
	& $ \mathbf{< 5.3 \times 10^{-5} } $	& CLEO \cite{Ammar:2001gi} 		 		& $ 6.74 \times 10^{-13} $ & $ > 1.1 \times 10^8 |V^d_{32}| $ \\
	(B^0 - K^0 \nu \bar{\nu})
	& $ (< 1.3 \times 10^{-5} ) $	 		& Belle \cite{Grygier:2017tzo} 			& & \\
\midrule
	D^\pm - \pi^\pm a 
	& $<1$ & & $ 1.11 \times 10^{-13} $ & $ > 3.3 \times 10^{5} |V^u_{21}| $ \\
	D^0 - \pi^0 a 
	& $<1$ & & $ 4.33 \times 10^{-14} $ & $ > 2.1 \times 10^{5} |V^u_{21}| $ \\
	D_s^\pm - K^\pm a
	& $<1$ & & $ 4.38 \times 10^{-14} $ & $ > 2.1 \times 10^{5} |V^u_{21}| $ \\
	B^0_s - \overbar{K}^0 a 
	& $<1$ & & $ 3.64 \times 10^{-13} $ & $ > 6.0 \times 10^{5} |V^d_{31}| $ \\
\bottomrule
\end{tabular}
\caption{
	Branching ratios (upper limits) and corresponding bounds (lower limits) on the PQ-breaking scale $ v_{PQ} $ from flavour-violating meson decays. 
	\textbf{Bold} font marks the current best limit from searches for $ P \to {P^\prime} a $, while parentheses mark the bound on the rare decay $ P \to {P^\prime} \nu \bar{\nu} $, which should be comparable. 
	Asterisks $ (^\ast) $ mark the expected reach of current or planned experiments.
}
\label{tab:2bodydecays}
\end{table}

\section{Axion-meson mixing}
\label{sec:mesonmixing}

In this section we discuss the mixing between axions and neutral hadronic mesons, and the impact on the meson oscillation probabilities.
Such a mixing effect can also lead to new contributions to both meson decays into axions and axion decays into two photons.
Although the mixing effect will turn out to be negligible for PQ axions which solve the strong $CP$ problem, it may be relevant for non-standard axions such as ALPs. 
Readers who are not interested in ALPs may skip this section, since it will not lead to any competitive bounds on PQ axions. 

\subsection{Parametrisation of mixing}

Axion-quark couplings in the mass-diagonal basis were discussed in Section~\ref{sec:physicalaxionbasis}.
Relevant to meson mixing are the terms
\begin{equation}
	\mathcal{L}'_\partial \supset
		- \frac{\partial_\mu a}{v_{PQ}} 
		\left[
			\sum_{q=u,d,s} c_q \bar{q} \gamma^\mu \gamma_5 q 
      		+ c_{sd} \bar{s} \gamma^\mu \gamma_5 d 
    		+ c_{sd}^\ast \bar{d} \gamma^\mu \gamma_5 s
     	\right] ,
\label{eq:aqq}
\end{equation}
where again
$ c_u = A^u_{11} + N_{DW} \beta_u/2$, 
$ c_d = A^d_{11} + N_{DW} \beta_d/2$, 
$ c_s = A^d_{22} + N_{DW} \beta_s/2$,
and 
$ c_{sd} = A^d_{21}$. 
These derivative couplings translate into effective axion-meson couplings
\begin{equation}
	\mathcal{L}^\mathrm{eff}_{aP} =
	- \sum_{P} c_P \frac{f_P}{v_{PQ}} 
   \partial_\mu a \partial^\mu P,
\label{eq:aP}
\end{equation}
where $f_P$ is the meson decay constant for $ P = \pi^0, \eta, \eta^\prime, K^0, \overbar{K}^0 $, and
$c_{\pi^0} = c_u-c_d $, 
$c_\eta = c_u + c_d - 2c_s$, 
$c_{\eta^\prime} = c_u + c_d + c_s $, and
$c_{K^0} = c_{sd} = c_{\overbar{K}^0}^\ast$.
This kinetic mixing can be diagonalised by the transformations
\begin{equation}
	a \to \frac{a}{\sqrt{1-\sum_P \eta_P^2}} , \qquad 
	P \to P + \frac{\eta_P a}{\sqrt{1-\sum_P \eta_P^2}} ,
\label{eq:kinmix}
\end{equation}
where $\eta_P \equiv c_P f_P/v_{PQ}$. 
This is naturally generalised to include also mesons containing $ c $ and $ b $ quarks.
For a QCD axion with $m_a \ll m_P$ and $f_a \gg f_P$, there is almost no impact on the standard meson dynamics. 
However, the results are valid for generalised ALPs, where the effect may be detectable.

\subsection{Meson mass splitting}

Axions and ALPs with off-diagonal quark couplings will mediate mixing between a heavy neutral meson $ P^0 $ ($ P = K $, $ D $, $ B $, or $ B_s $) and its antiparticle $ \overbar{P}^0 $ in addition to that from weak interactions.
An explicit calculation, showing how axion interactions yield an additional contribution to meson mass splittings, is given in Appendix~\ref{app:sec:mesonmixing}.
We quote the result, namely that 
\begin{equation}
	(\Delta m_P)_\mathrm{axion}
		\simeq |\eta_P|^2 m_P
		= |c_P|^2 \frac{f_{P^0}^2}{v^2_{PQ}} m_P .
\end{equation}
The total mass difference is then given by $ \Delta m_P = (\Delta m_P)_\mathrm{SM} + (\Delta m_P)_\mathrm{axion} $.
As an example, consider the effect of axion-kaon mixing on the $ K^0_L - K^0_S $ mass difference, experimentally measured to be 
$(\Delta m_K)_\mathrm{exp} = (3.484 \pm 0.006) \times 10^{-12} $ MeV \cite{Patrignani:2016xqp}. 
The error is dominated by the theory uncertainty, which may be large \cite{Brod:2011ty}; near-future lattice calculations aim to reduce the error on $ \Delta m_K $ to $ \mathcal{O}(20\%) $ \cite{Bai:2018mdv}, with further improvements from next-generation machines. 
As a conservative estimate, we shall only demand the axion contribution to any $ \Delta m_P $ is not larger than the experimental central value.
We then have
$|\eta_{K^0}| \simlt 8 \times 10^{-8} $, which (assuming $ c_{K^0} \approx 1 $) corresponds to the bound $ v_{PQ} \simgt 2 \times 10^6 $ GeV.
Similar results for $ D $, $ B $ and $ B_s $ mixing are tabulated in Table~\ref{tab:massdifferences}. 
Belle-II is expected to improve the sensitivity of $ D^0 - \overbar{D}^0 $ mixing by about one order of magnitude with the full 50 ab$^{-1}$ of data \cite{Li:2018vjr}.

\begin{table}[!ht]
\centering
\begin{tabular}{lrr}
\toprule
	System & 
	$ (\Delta m_P)_\mathrm{exp} /\mathrm{MeV} $ & 
	$ v_{PQ} /\mathrm{GeV} $ \\
\midrule
$K^0 - \overbar{K}^0$ & $(3.484 \pm 0.006) \times 10^{-12}$ & $ \simgt 2 \times 10^{6} |c_{K^0}| $ \\
$D^0 - \overbar{D}^0$ & $(6.25 \,^{+2.70}_{-2.90}) \times 10^{-12}$ & $ \simgt 4 \times 10^{6} |c_{D^0}| $ \\
$B^0 - \overbar{B}^0$ & $(3.333 \pm 0.013) \times 10^{-10}$ & $ \simgt 8 \times 10^{5} |c_{B^0}| $ \\
$B_s^0 - \overbar{B}_s^0$ & $(1.1688 \pm 0.0014) \times 10^{-8}$ & $ \simgt 1 \times 10^{5} |c_{B^0_s}|$ \\
\bottomrule
\end{tabular}
\caption{%
	Limits on $ v_{PQ} $ from contributions to neutral meson mass differences.
	Measured values of $ \Delta m_P $ are given in the PDG \cite{Patrignani:2016xqp}.
	Meson decay constants $ f_{P^0} $ are extracted from global averages given in \cite{Rosner:2015wva}.
}
\label{tab:massdifferences}
\end{table}

\subsection{Axion-pion mixing and ALPs}

We have seen that axion-meson kinetic mixing can affect the oscillation probability (and thereby the mass difference) of neutral heavy mesons, arising from off-diagonal quark couplings of axions. 
In this subsection, we will see that even flavour-diagonal couplings can lead to interesting consequences. 
As shown in Eqs.~\ref{eq:aqq} and \ref{eq:aP}, there arises in particular axion-pion kinetic mixing as a consequence of the physical $ \pi^0 $ containing a small admixture of the nominal axion and \emph{vice versa}. 
This induces axion contributions to any process normally involving $ \pi^0 $.

Kinetic diagonalisation (as in Eq.~\ref{eq:kinmix}) induces mass couplings of the form $ -\tfrac{1}{2} \Phi^T M^2_\Phi \Phi $, where $ \Phi = (a, \pi^0) $ and 
\begin{equation}
	M^2_\Phi = m_{\pi^0}^2 \pmatr{
		\dfrac{m_a^2}{m_{\pi^0}^2} + \dfrac{\eta_{\pi^0}^2}{1-\eta_{\pi^0}^2} & \dfrac{\eta_{\pi^0}}{\sqrt{1-\eta_{\pi^0}^2}} \\
		\dfrac{\eta_{\pi^0}}{\sqrt{1-\eta_{\pi^0}^2}} & 1 \\
	} ,
\label{eq:pionmixingmassmatrix}
\end{equation}
where $\eta_{\pi^0} = c_{\pi^0} f_{\pi}/v_{PQ} = \tilde  c_{\pi^0} f_{\pi}/f_a $, with $\tilde  c_{\pi^0} \equiv  c_{\pi^0}/N_{DW}$.
This is subsequently diagonalised by a $ 2\times 2 $ rotation in terms of an angle $ \theta_\pi $, where
\begin{equation}
	 \tan 2\theta_\pi = 
	 \frac{2 m^2 _{\pi^0} \eta_{\pi^0} \sqrt{1-\eta_{\pi^0}^2}}{m^2 _{\pi^0}(1-2 \eta^2_{\pi^0})-m_a^2(1-\eta_{\pi^0}^2)} .
\end{equation}
Starting from the canonical physical basis in Eq.~\ref{eq:aqq}, the physical basis accounting also for kinetic mixing is thus obtained by field transformations
\begin{equation}
\begin{split}
	a &\to 
	\frac{\cos{\theta_{\pi}}\, a + \sin{\theta_\pi}\, \pi^0}{\sqrt{1-\eta_{\pi^0}^2}} ,
	\\
	\pi^0 &\to 
	\left(\cos{\theta_\pi} + \frac{\sin{\theta_\pi} \eta_{\pi^0}}{\sqrt{1-\eta_{\pi^0}^2}} \right) \, \pi^0 
	- \left( \sin{\theta_\pi} - \frac{\cos{\theta_\pi} \eta_{\pi^0}}{\sqrt{1-\eta_{\pi^0}^2}} \right) \, a .
\end{split}
\end{equation}
To leading order in $ \eta_{\pi^0} $, we have
\begin{equation}
	a \to 
	a + \frac{\eta_{\pi^0} m_{\pi^0}^2}{m_{\pi^0}^2 - m_a^2} \pi^0 ,
	\qquad
	\pi^0 \to 
	\pi^0 - \frac{\eta_{\pi^0} m_a^2}{m_{\pi^0}^2 - m_a^2} a .
\label{eq:a-pi}
\end{equation}
For a QCD axion with $ m_a \ll m_\pi^0 $ and $\eta_{\pi^0}\ll 1$, its contribution to the physical pion is vanishingly small.
However, this mixing may be interesting for more general ALPs, where the mass and decay constant are not necessarily correlated.

The axion-meson mixing effect discussed above can modify decays of heavy mesons to lighter mesons plus an axion, as well as to the decay of an axion to two photons. 
The basic idea is very simple: in the standard hadronic decay of a heavy meson into two pions, one of the neutral pions in the final state can convert into an axion via the mixing effect discussed above, leading to a final state consisting of an axion. 
Similarly, the standard decay of a neutral pion into two photons can also mediate the decay on an axion into two photons.

Applying Eq.~\ref{eq:a-pi} to an ALP, still denoted by $a$, perhaps the most interesting processes induced by mixing are $ K^+ \to \pi^+ a $ and $ a \to \gamma\gamma $.
Considering only the mixing-induced effect, we have 
\begin{equation}
\Gamma(K^+ \to \pi^+ a) \simeq 
		\left( 
			\frac{\eta_{\pi^0} m_a^2}{m_{\pi^0}^2 - m^2_a} 
		\right)^2 
		\Gamma(K^+ \to \pi^+ \pi^0)  .
\label{eq:Kpia}
\end{equation} 
Taking the ballpark of $\mathrm{Br}(K^+\to \pi^+ a) \simlt 10^{-10}$ listed in Table 2 and  $ \mathrm{Br}(K^+ \to \pi^+ \pi^0) = 20.67\%  $, we find a mass-dependent bound
\begin{equation}
	f_a \simgt 4 \left(
	\frac{ \tilde c_{\pi^0} m_a^2}{m^2_{\pi^0} -m_a^2}\right)  \mathrm{TeV} 
\label{eq:mixKpi}
\end{equation}
which is applicable for $m_a \simlt 110$ MeV. 
Similarly, one finds the axion decay to photons
\begin{equation}
\Gamma(a \to \gamma\gamma) \simeq 
		\left( 
			\frac{\eta_{\pi^0} m_a^2}{m_{\pi^0}^2 - m^2_a} 
		\right)^2 
		\left(
			\frac{m_a}{m_{\pi^0}}
		\right)^3 
		\Gamma(\pi^0 \to \gamma\gamma) .
\label{eq:agg}
\end{equation}
In the SM with massless valence quarks and $ N_C = 3 $ colours, we have \cite{Miskimen:2011zz}
\begin{equation}
	\Gamma(\pi^0 \to \gamma\gamma) = 
	\frac{\alpha^2 m_{\pi^0}^3 N_C^2}{576 \pi^3 f_\pi^2} 
	\simeq 7.63 \mathrm{~eV}.
\label{eq:pigg}
\end{equation}
The standard form of the axion-photon coupling, $ \frac{1}{4} g_{a\gamma} a F \tilde{F}$, gives $\Gamma(a\to\gamma\gamma) = \frac{1}{64\pi} g_{a\gamma}^2 m_a^3$.
We may then write the mixing-induced axion-photon coupling as
\begin{equation}
	(g_{a\gamma})_\mathrm{mix} \simeq 
	\frac{\alpha}{\pi} \frac{\tilde c_{\pi^0} m_a^2}{m^2_{\pi^0} -m_a^2} 
	\frac{1}{f_a} .
\label{eq:mgag}
\end{equation}
Therefore the bound in Eq.~\ref{eq:mixKpi} corresponds to
\begin{equation}
	(g_{a\gamma})_\mathrm{mix} \lesssim 5.8\times10^{-7} \, \mathrm{GeV}^{-1}
 ~~\mbox{for}~~m_a \lesssim 110\, \mbox{MeV}. 
\label{eq:mgagK}
\end{equation}
Extensive studies of ALPs over a wide range of parameter space (summarised in e.g. Fig.~1 of \cite{Jaeckel:2017tud}) place very strong bounds for $ g_{a\gamma} < 10^{-10} \, \mathrm{GeV}^{-1} $ for the whole range of $ m_a \simlt 100 $ MeV, which translates to
\begin{equation}
	f_a \simgt 2\times 10^7 \left(\frac{\tilde c_{\pi^0} m_a^2}{m^2_{\pi^0} -m_a^2}\right)
\left( \frac{10^{-10} \, \mathrm{GeV}^{-1}}{g_{a\gamma}} \right) \mbox{GeV}.
\end{equation}
Let us finally note that the E787 experiment searched for $ K^+\to \pi^+ a $ followed by $ a\to \gamma\gamma $ in the range of $ m_a = 5-100 $ MeV \cite{Kitching:1997zj}. 
Combining the two expressions in Eqs.~\ref{eq:Kpia} and \ref{eq:mgag}, the E787 result gives (for $ m_a = 10-96 $ MeV) the bound 
\begin{equation}
	(g_{a\gamma})_\mathrm{mix} \simlt 5 \times 10^{-5} \, \mathrm{GeV}^{-1} ,
\end{equation}
which is less stringent than Eq.~\ref{eq:mgagK}.

\section{Lepton decays}
\label{sec:leptondecays} 

\subsubsection*{\boldmath{$\ell_1 \to \ell_2 a$}}

Two-body lepton decays of the form $ \ell_1 \to \ell_2 a $ follow analogously to meson decays, with the notable difference that both axial and vector couplings contribute, since the decaying particle has non-zero spin. 
We define a total coupling $ C^e_{\ell_1 \ell_2} $ by
\begin{equation}
	\left| C^e_{\ell_1 \ell_2} \right|^2 = \left| V^e_{\ell_1 \ell_2} \right|^2 + \left| A^e_{\ell_1 \ell_2} \right|^2 .
\end{equation}
As done for mesons in Eqs.~\ref{eq:BrPYcPY}--\ref{eq:cPY}, the branching ratio may once again be written in terms of a coefficient $ \tilde{c}_{\ell_1 \to \ell_2} $, by
\begin{equation}
	\mathrm{Br}(\ell_1 \to \ell_2 a)
	= \tilde{c}_{\ell_1 \to \ell_2} \left| C^e_{\ell_1 \ell_2} \right|^2
		\left(\frac{10^{12} \mathrm{~GeV}}{v_{PQ}}\right)^2 ,
\end{equation}
where
\begin{equation}
	\tilde{c}_{\ell_1 \to \ell_2} = 
		\frac{1}{16\pi \, \Gamma(\ell_1)}
		\frac{m_{\ell_1}^3}{(10^{12} \mathrm{~GeV})^2}
		\left( 1 - \frac{m_{\ell_2}^2}{m_{\ell_1}^2} \right)^3 .
\end{equation}
These are evaluated, with corresponding limits placed on $ v_{PQ} $, for the three possible lepton decays.
The results are tabulated in Table~\ref{tab:2bodydecaysleptons}.

The most interesting of these is $ \mu^+ \to e^+ a $, for which the SM background consists almost entirely of ordinary $ \beta $ decay, $ \mu^+ \to e^+ \nu \bar{\nu} $. 
The muon decay width $ \Gamma_\mu $ is given to good approximation by $ \Gamma_\mu \simeq \Gamma(\mu^+ \to e^+ \nu \bar{\nu}) \simeq G_F^2 m_\mu^5 / (192 \pi^3) $.
Assuming $ \mu^+ \to e^+ a $ decays are isotropic, i.e. the decay is purely vectorial (or axial), the experiment at TRIUMF provides the limit $ \mathrm{Br}(\mu^+ \to e^+ a) < 2.6 \times 10^{-6} $ \cite{Jodidio:1986mz}, corresponding to $ v_{PQ}/|V^e_{21}| ~(\mathrm{or~} |A^e_{21}|)> 5.5 \times 10^9 $ GeV.
They searched specifically for decays with an angular acceptance $ \cos \theta > 0.975 $, where $ \theta $ is the positron emission angle; in this region SM three-body decays are strongly suppressed.
The TWIST experiment \cite{Bayes:2014lxz} has performed a broader search, accommodating non-zero anisotropy $ A $ as well as massive bosons, but are less sensitive for isotropic decays in the massless limit.
The limits for isotropic ($ A = 0 $) and maximally anisotropic ($ A = \pm 1 $) decays are given in Table~\ref{tab:2bodydecaysleptons}.

Let us sketch the angular dependence of $ \mu \to e a $ decays, which are not generally isotropic, as these would relate to TWIST; the formulas generalise immediately to $ \tau $ decays.
Consider $ \mu^+ $ with a polarisation $ \eta = (0,\boldsymbol{\eta}) $ decaying into a positron with helicity $ \lambda_e = \pm 1 $ and momentum $ k_e $, as well as an axion.
Neglecting $ m_e $ and $ m_a $,
\begin{equation}
	\left| \mathcal{M} \right|^2 = 
	\frac{m_{\mu}^3}{v_{PQ}^2} 
	\sum_{\lambda_e = \pm 1}
		\Big[
			|C^e_{21}|^2
			\left( m_{\mu} - 2 \lambda_e (\eta \cdot k_e) \right)
			+ 2 \mathrm{Re}[A^e_{21} (V^e_{21})^\ast] \left( m_{\mu} - 2 (\eta \cdot k_e) \right)
		\Big],
\end{equation}
where $ \eta \cdot k_e = - |\mathbf{k}_e| \cos \vartheta_{\eta e} $.
We can describe the degree of muon polarisation $ P_\mu $ as the projection of $ \boldsymbol{\eta} $ onto the beam direction $ \mathbf{\hat{z}} $, i.e.
$ P_\mu \simeq \cos \vartheta_{\eta z} = {\boldsymbol{\eta} \cdot \mathbf{\hat{z}}}/{|\boldsymbol{\eta}|} $.
For a more precise treatment one should consider the distribution of $ \boldsymbol{\eta} $ in a muon ensemble, but as we shall assume all muons are highly polarised opposite to the beam direction, i.e. $ P_\mu \sim -1 $, this is sufficient for our purposes.
TWIST measures the positron emission angle $ \theta = \vartheta_{\eta z} - \vartheta_{\eta e} $; for highly polarised muons, we have $ \cos \vartheta_{\eta e} \simeq P_\mu \cos \theta $.
Summing over $ \lambda_e $, the differential decay rate is given by
\begin{equation}
	\frac{\mathrm{d} \Gamma}{\mathrm{d} \cos \theta} 
	= \frac{\overbar{\left| \mathcal{M} \right|^2} }{32 \pi m_\mu} 
	\simeq
	\frac{|C^e_{21}|^2}{32 \pi} 
	\frac{m_\mu^3}{v_{PQ}^2} 
		( 1 - A P_\mu \cos \theta ),
\end{equation}
where we define the anisotropy
\begin{equation}
	A =- \frac{2 \mathrm{Re}[A^e_{21} (V^e_{21})^\ast]}{|C^e_{21}|^2} .
\label{eq:anisotropy}
\end{equation}
The limiting cases are $ A^e_{21} = V^e_{21} $, giving $ A = - 1 $ (corresponding to an SM-like $ V-A $ current interaction), or $ A^e_{21} = -V^e_{21} $, giving $ A = 1 $ (a $ V+A $ interaction).
The signal strength with respect to the SM background is maximised for $ A = 1 $, particularly in the region with $ \cos \theta \sim 1 $.
The A to Z model, discussed below, predicts exactly this scenario, although the high predicted PQ scale $ v_{PQ} \sim 10^{12} $ GeV implies the signal is very small despite the enhancement.

Finally, the Mu3e experiment, primarily designed to look for $ \mu \to eee $ (discussed below), can also be used to test for $ \mu \to e a $, and tentatively probe scales of $ v_{PQ} \simgt 10^{10} $ GeV \cite{PerrevoortThesis} by the end of its run.

\begin{table}[!ht]
\centering
\begin{adjustbox}{width=1.0\textwidth,center=\textwidth}
\begin{tabular}{D{@}{\ \to \ }{3,3} rrrrr}
\toprule
	\multicolumn{1}{c}{Decay} & 
	Branching ratio & 
	Experiment &
	$ \tilde{c}_{\ell_1 \to \ell_2} $ &
	$ v_{PQ} /\mathrm{GeV} $ \\
\midrule
	\mu^+ @ e^+ a 				
	& $ < 2.6 \times 10^{-6} $	& $ (A=0) $ Jodidio \emph{et al} \cite{Jodidio:1986mz}	& $ 7.82 \times 10^{-11}$ 	& $ > 5.5 \times 10^9 |V^e_{21}| $ \\
	& $ < 2.1 \times 10^{-5}  $ 		& $ (A=0) $ TWIST \cite{Bayes:2014lxz} 		 	&  							& $ > 1.9 \times 10^9 |C^e_{21}| $ \\
	& $ < 1.0 \times 10^{-5} $ 	& $ (A=1) $ TWIST \cite{Bayes:2014lxz}			&  							& $ > 2.8 \times 10^9 |C^e_{21}| $ \\
	& $ < 5.8 \times 10^{-5} $ 	& $ (A=-1) $ TWIST \cite{Bayes:2014lxz}			&  							& $ > 1.2 \times 10^9 |C^e_{21}| $ \\
	& $ \simlt 5 \times 10^{-9} $*		& Mu3e (future) \cite{PerrevoortThesis} 		& 							& $ \simgt 1 \times 10^{11} |C^e_{21}| $\\
	\tau^+ @ e^+ a 				
	& $ < 1.5 \times 10^{-2} $ 			& ARGUS \cite{Albrecht:1995ht} 					& $ 4.92 \times 10^{-14}$ 	& $ > 1.8 \times 10^6 |C^e_{31}| $ \\
	\tau^+ @ \mu^+ a 			
	& $ < 2.6 \times 10^{-2} $ 			& ARGUS \cite{Albrecht:1995ht} 					& $ 4.87 \times 10^{-14}$ 	& $ > 1.4 \times 10^6 |C^e_{32}| $ \\
\bottomrule
\end{tabular}
\end{adjustbox}
\caption{%
	Branching ratios (upper limits) and corresponding bounds (lower limits) on $ v_{PQ} $ from two-body cLFV decays. 
	The assumed anisotropy $ A $ can be related to the formula in Eq.~\ref{eq:anisotropy}.
}
\label{tab:2bodydecaysleptons}
\end{table}

\subsubsection*{\boldmath{$\ell_1 \to \ell_2 a \gamma$}} 

Additionally, we may examine decays with an associated photon, i.e. 
$ \ell_1 \to \ell_2 a \gamma $.
These can be studied in experiments searching for $ \ell_1 \to \ell_2 \gamma $, which, if experimentally measured, are unequivocal signs of new physics; in the SM, $ \mathrm{Br}(\mu \to e \gamma) \sim 10^{-54} $, certainly unobservable.
The differential decay rate for $ \ell_1 \to \ell_2 a \gamma $ in the limit of $ m_{\ell_2} = m_a = 0 $ may be expressed by 
\begin{equation}
	\frac{\mathrm{d}^2 \Gamma}{\mathrm{d}x \, \mathrm{d}y} 
	= \frac{\alpha \left| C^e_{\ell_1 \ell_2} \right|^2 m_{\ell_1}^3}{32 \pi^2 v_{PQ}^2} f(x,y), 
	\qquad
	f(x,y) = \frac{(1-x)(2-y-xy)}{y^2(x+y-1)},
\end{equation}
where $ f(x,y) $ is a function of 
$ x = 2E_{\ell_2}/m_{\ell_1} $, 
$ y = 2E_\gamma/m_{\ell_1} $, i.e. (twice) the fraction of invariant mass carried away by the lighter lepton and photon, respectively.
Energy conservation requires $ x, y \leq 1 $ and $ x + y \geq 1 $.
Moreover, the angle $ \theta_{2\gamma} $ between $ \ell_2 $ and the photon is fixed by kinematics to
\begin{equation}
	\cos \theta_{2\gamma} = 1 + \frac{2(1-x-y)}{xy}.
\end{equation}
Alternatively one can write the decay rate in terms of $ x $ and $ c_\theta \equiv \cos \theta_{2\gamma} $, i.e.
\begin{equation}
	\frac{\mathrm{d}^2 \Gamma}{\mathrm{d}x \, \mathrm{d} c_\theta} 
	= \frac{\alpha \left| C^e_{\ell_1 \ell_2} \right|^2 m_{\ell_1}^3}{32 \pi^2 v_{PQ}^2} f(x,c_\theta), 
	\qquad
	f(x,c_\theta) = \frac{1-x(1-c_\theta)+x^2}{(1-x)(1-c_\theta)}.
\end{equation}
We may relate the branching ratios of decays with and without a radiated photon by
\begin{equation}
	\mathcal{R}_{\ell_1\ell_2}
	= \frac{\mathrm{Br}(\ell_1 \to \ell_2 a \gamma)}{\mathrm{Br}(\ell_1 \to \ell_2 a)} 
	= \frac{\alpha}{2\pi}
	\int \mathrm{d}x \, \mathrm{d}y f(x,y) .
\end{equation}

The radiative decay possesses two divergences: an IR divergence due to soft photons ($ x \simeq 1 $) and a collinear divergence ($ \theta_{2\gamma} \simeq 0 $). 
In practice, appropriate cuts are made on the minimum photon energy and angular acceptance well away from the IR-divergent region.
Such cuts were discussed in the context of $ \ell_1 \to \ell_2 \gamma $ decays \cite{Hirsch:2009ee}, in particular as they related to LAMPF \cite{Goldman:1987hy} and MEG \cite{TheMEG:2016wtm} experiments.
The region of interest for MEG is for $ x, y \simeq 1 $, or equivalently $ c_\theta \simeq \pi $, where the SM background disappears.
However, decays with an associated flavoured axion are also suppressed in this limit, i.e. the integral $ \int f $ vanishes for very soft axions.
One might consider a broader region of phase space, provided the induced backgrounds%
\footnote{
	The primary sources of background are radiative muon decay (RMD) and accidental $ e^+ e^- $ annihilation-in-flight (AIF). 
	For large photon energies and increased stopped muon rate, AIF dominates over RMD.
} 
are under control.
A comprehensive experimental study of such signals, e.g. for the MEG-II upgrade  \cite{Baldini:2018nnn}, would be welcome.
An explicit limit on $ \mu \to e f \gamma $, where $ f $ is a light scalar or pseudoscalar, is given by the Crystal Box experiment, which sets $ \mathrm{Br}(\mu \to e f \gamma) < 1.1 \times 10^{-9} $ at 90\% CL \cite{Bolton:1988af}.
Unlike the TRIUMF experiment \cite{Jodidio:1986mz} discussed above, this limit does not assume isotropic decays.
Using the same cuts%
\footnote{The Crystal Box analysis uses the cuts $ E_\gamma, E_e > 38 $ MeV, $ \theta_{e\gamma} > 140^\circ $.}
we find $ \int f \simeq 0.011 $, yielding the bound $ v_{PQ}/\mathrm{GeV} > 9.4 \times 10^8 |C^e_{21}| $.
In Table~\ref{tab:3bodydecaysleptons} we summarise current and future experimental limits on branching ratios of $ \ell_1 \to \ell_2 \gamma $.

\begin{table}[!ht]
\centering
\begin{tabular}{D{@}{\ \to \ }{3,3} rr}
\toprule
	\multicolumn{1}{c}{Decay} & 
	Branching ratio & 
	Experiment \\
\midrule
	\mu^+ @ e^+ \gamma	& $ < 4.2 \times 10^{-13} $ 	& MEG \cite{TheMEG:2016wtm}  \\
						& $ \simlt 6 \times 10^{-14} $*	& MEG-II (future) \cite{Baldini:2018nnn} \\
	\tau^{-} @ e^{-} \gamma	& $ < 3.3 \times 10^{-8} $ & BaBar \cite{Aubert:2009ag} \\ 
	\tau^{-} @ \mu^{-} \gamma	& $ < 4.4 \times 10^{-8} $ & BaBar \cite{Aubert:2009ag} \\ 
\bottomrule
\end{tabular}
\caption{%
Experimental upper limits on cLFV decays $ \ell_1 \to \ell_2 \gamma $.
}
\label{tab:3bodydecaysleptons}
\end{table}

Also radiative $ \beta $ decay itself, $ \mu \to e \nu \bar{\nu} \gamma $, can give information on decays to axions.
The most precise measurement comes from MEG, giving $ \mathrm{Br}(\mu \to e \nu \bar{\nu} \gamma) = (6.03 \pm 0.14 (\mathrm{stat}) \pm 0.53 (\mathrm{sys})) \times 10^{-8} $ for $ E_e > 45 $ MeV and $ E_\gamma > 40 $ MeV, in agreement with the SM \cite{Adam:2013gfn}.
Requiring the axion decay to not significantly exceed the error on this measurement, i.e. $ \mathrm{Br}(\mu \to e a \gamma) \simlt 1 \times 10^{-8} $, yields a limit $ v_{PQ}/|C^e_{\mu e}| \simgt 1.2 \times 10^8 $ GeV.
We see that the limit from $ \mu \to e a $ is stronger by approximately a factor 40.

\subsubsection*{\boldmath{$\mu \to eee$} and \boldmath{$\mu - e$} conversion}

We may also consider processes without an axion in the final state.
Axion mediation will induce the decay $ \mu \to eee $, although the presence of two axion vertices and additional suppression by $ 1/v_{PQ} $ means these processes are again only interesting for ALPs.
The current upper bound on the branching ratio is $ \mathrm{Br}(\mu^+ \to e^+ e^- e^+) < 1.0 \times 10^{-12} $, set by SINDRUM \cite{Bellgardt:1987du}.
The Mu3e experiment \cite{Blondel:2013ia} currently under development is expected to start taking data in 2019, and will significantly improve the sensitivity by four orders of magnitude, i.e. $ \mathrm{Br}(\mu \to eee) \simlt 1 \times 10^{-16} $.
To lowest order in $ m_e^2 $, the branching ratio for the axion-mediated decay is given by
\begin{equation}
\begin{split}
	\mathrm{Br}(\mu^+ \to e^+ e^- e^+) 
	&\approx 
	\frac{m_e^2 m_\mu^3}{16 \pi^3 \Gamma(\mu)} 
	\frac{|A^e_{11}|^2 |C^e_{21}|^2}{v_{PQ}^4}
	\left( \ln \frac{m_\mu^2}{m_e^2} - \frac{15}{4} \right) , \\
	&\approx
	1.43 \times 10^{-41}
	|A^e_{11}|^2 |C^e_{21}|^2
	\left(\frac{10^{12} \mathrm{~GeV}}{v_{PQ}}\right)^4 .
\end{split}
\end{equation}
Assuming $ \mathcal{O}(1) $ couplings, we see that such decays are only reachable by experiment provided $ v_{PQ} \simlt 10^6$ GeV.

As the axion (or ALP) also couples to quarks, one may consider $ \mu - e $ conversion in nuclei, mediated by the axion.
The relevant couplings are now $ C^e_{21} $ and the axion-nucleon coupling $ g_{aN} = C_{aN} m_N/v_{PQ} $.
The numerical factor $ C_{aN} $ is model-dependent, given in terms of flavour-diagonal couplings of the up and down quarks.
In standard cases these are essentially given by the quark PQ charges (see e.g. \cite{Srednicki:1985xd} for standard formulae), but in more general scenarios such as a flavoured axion, these can deviate significantly.%
\footnote{
	It is even possible to suppress the nucleon couplings entirely, yielding a \emph{nucleophobic} axion \cite{DiLuzio:2017ogq}.
}
The axion-mediated $\mu-e$ conversion is a spin-dependent process which was discussed in  \cite{Cirigliano:2017azj}. The conversion-to-capture ratio in a nucleus $(A,Z)$ is qualitatively given by
\begin{equation}
 R_{\mu e}^{(A,Z)} \equiv  
 \frac{\Gamma({\mu^- \to e^- (A,Z)})}{\Gamma^{(A,Z)}_{\mu^- \mathrm{cap}}}
	\sim 
 \frac{m^5_\mu}{(q^2-m_a^2)^2}
  \frac{(\alpha Z)^3}{\pi^2\, \Gamma^{(A,Z)}_{\mu^- \rm{cap}}}
 \frac{ m_\mu^2 m_N^2}{v_{PQ}^4}  |C^e_{21}|^2 |S^{(A,Z)}_N C_{aN}|^2 ,
\end{equation}
where  $q^2 \approx m_\mu^2$ is the momentum-transfer  and $S_N^{(A,Z)}$ is the 
total nucleon spin of a nucleus $(A,Z)$. Not accounted for here are nuclear spin and structure form factors, which were discussed in \cite{Cirigliano:2017azj} and are $ \mathcal{O}(1) $.
The suppression by $ v_{PQ}^4 $ suggests $ \mu - e $ conversion is only realistically detectable in ALP scenarios. 
The current best limit comes from SINDRUM-II: $ R^\mathrm{Au}_{\mu e}  < 7 \times 10^{-13} $ \cite{Bertl:2006up}. 
Assuming again $ \mathcal{O}(1) $ couplings and form factors, SINDRUM-II sets 
$v_{PQ} \simgt 10^6 $ GeV, comparable to the $\mu \to 3e$ bound. 
The upcoming experiments Mu2e and COMET are both looking for $ \mu^- \mathrm{Al} \to e^- \mathrm{Al} $, and both aim to probe $ R_{\mu e} < 6 \times 10^{-17} $ at 90\% CL \cite{Donghia:2018duf,Wu:2017zwh}, a factor $ 10^4 $ improvement over the SINDRUM result.

\section{A to Z Pati-Salam Model}
\label{sec:AtoZ}

We present here a recently proposed QCD axion model \cite{Bjorkeroth:2017tsz}, based on the rather successful A to Z model \cite{King:2014iia}, which seeks to resolve the flavour puzzle by way of Pati-Salam unification coupled to an $ A_4 \times \mathbb{Z}_5 $ family symmetry.
The family symmetry is completely broken by gauge singlet flavons $ \phi $, which are triplets under $ A_4 $ and couple to left-handed SM fields.
However, information about the underlying symmetry remains in the particular vacuum structure of the flavons.
The initial viability of the model, which predicts certain Yukawa structures based on the so-called CSD(4) vacuum alignment, was demonstrated in \cite{King:2014iia},
and leptogenesis was considered in \cite{DiBari:2015oca}.

In \cite{Bjorkeroth:2017tsz}, we updated and improved the numerical fit to flavour data, as well as demonstrating that, with small adjustments, the A to Z model can resolve the strong $ CP $ problem.
The axion then emerges from the same flavons that are responsible for SM Yukawa couplings; in other words, no additional field content is necessary to realise a PQ axion.
Moreover, as all Yukawa couplings are fixed by the fit to data, also the axion couplings are known exactly, with no additional free parameters.
As the focus of this work is on axion couplings to matter, we limit our discussion primarily to the resultant Yukawa and mass matrices of the SM fermions.
However in Appendix~\ref{app:sec:couplings} we derive explicitly the axion-matter couplings from the Yukawa superpotential.
In Appendix~\ref{app:sec:numericalfit} we provide the best fit parameters for the A to Z model and corresponding axion couplings.

\subsection{Mass matrices and parameters}

The charged fermion Yukawa matrices are given at the GUT scale by
\begin{equation}
\begin{split}
	Y^u = \pmatr{
		0 & b & \epsilon_{13} c \\
		a  & 4 b & \epsilon_{23} c \\
		a  & 2 b & c \\
	}, 
	\quad
	Y^d = \pmatr{
		y_d^0  & 0 & 0 \\
		B y_d^0  & y_s^0  & 0 \\
		B y_d^0  & 0 & y_b^0
	}, 
	\quad
	Y^e = \pmatr{
		-(y_d^0/3)  & 0 & 0 \\
		B y_d^0  & x y_s^0  & 0 \\
		B y_d^0  & 0 & y_b^0
	} .
\end{split}
\end{equation}
All parameters are dimensionless and in general complex, although three can be immediately made real by an overall rephasing of the three Yukawa matrices.
The mass matrix of the light Majorana neutrinos (after seesaw) is
\begin{equation}
	m^\nu = 
	m_a \pmatr{0&0&0\\0&1&1\\0&1&1}
	+ m_b e^{i \eta} \pmatr{1&4&2\\4&16&8\\2&8&4}
	+ m_c e^{i \xi} \pmatr{0&0&0\\0&0&0\\0&0&1},
\end{equation}
where $ m_i $ are real, with dimensions of mass and $ \eta $, $ \xi $ are phases.

Note that the scales of the various free parameters are constrained by the model itself. 
By rather simple assumptions about the flavon VEVs, discussed fully in \cite{King:2014iia}, and assuming all dimensionless couplings in the renormalisable theory are $ \mathcal{O}(1) $, we may infer generic properties of the parameters.
Parameters $ a $, $ b $ and $ c $ correspond closely to the three up-type quark Yukawa couplings, i.e. $ a \ll b \ll c \sim 1 $.
Meanwhile $ y_d^0 $, $ y_b^0 $ and $ y_s^0 $ are correlated with the down-type quark Yukawa couplings, i.e. $ y_d^0 \ll y_b^0 \ll y_s^0 $.
$ B $ is an $ \mathcal{O}(1) $ ratio of couplings, and $ \epsilon_{i3} \ll 1 $ are small perturbations of a flavon VEV.
The $ \mathcal{O}(1) $ factor $ x $ is a Clebsch-Gordan factor, introduced by additional Higgs multiplets in a variation of the Georgi-Jarlskog mechanism.
In the neutrino sector, the principle of sequential dominance on which the model relies demands a normal ordering and strong mass hierarchy, with $ m_a \gg m_b \gg m_c $, predicting the lightest neutrino with a mass of $ < 1 $ meV.
A fit of these parameters to data has been performed \cite{Bjorkeroth:2017tsz}, with central results collected in Appendix~\ref{app:sec:numericalfit}.
The model is fitted to experimental results%
\footnote{
	In \cite{Antusch:2013jca} they perform the running of low-scale experimental results (from global fits) up to the GUT scale, assuming the MSSM; they provide GUT-scale values for quark and charged lepton Yukawa couplings, and CKM mixing parameters.
}
by an MCMC analysis.
Bayesian credible intervals are also provided, showing that despite a large number of free parameters, small tensions in the predictions for $ \theta^\ell_{23} $ and $ \delta^\ell $ may be further probed by increased sensitivity in current and future neutrino experiments.

The PQ-breaking scale $ v_{PQ} $ is determined primarily by the largest VEV among the flavons $ \phi $ carrying PQ charge.
The VEV of this flavon (named $ \phi_2^u $) is proportional to the parameter $ b $ in $ Y^u $, which in turn is dominantly responsible for the charm quark Yukawa coupling; as the third generation largely does not couple to the PQ symmetry, this is the heaviest relevant fermion in the flavoured axion theory.
The numerical fit gives $ |b| = 3.4 \times 10^{-3} $.
The details of how the flavons and parameters are related are given in Appendix~\ref{app:sec:couplings}, showing that $ b \sim \braket{\phi_2^u}/M_\mathrm{GUT} \Rightarrow v_{PQ} \simeq \braket{\phi_2^u} \sim 10^{12} $ GeV.

\subsection{Predictions}

Once the fermion mixing matrices are known from the fit, we can immediately determine the vector and axial coupling matrices $ V^f $ and $ A^f $ using Eqs.~\ref{eq:VfAf}.
Recalling that $ V^f $ and $ A^f $ are Hermitian, we have
\begin{equation}
\begin{split}
	V^u = -A^u &\simeq \pmatr{
	1.0 							& 4.3 \times 10^{-3} e^{-0.05i} & -1.7 \times 10^{-5} e^{-0.015i} \\
	4.3 \times 10^{-3} e^{0.05i} 	& -0.5 							&  -6.0 \times 10^{-4} \\
	-1.7 \times 10^{-5} e^{0.015i} 	& -6.0 \times 10^{-4} 			& 7.3 \times 10^{-7}
	} , \\ 
	V^d = -A^d &\simeq \pmatr{
	0.78 	& 0.25		& -0.0065 \\
	0.25 	& 0.72 		& -0.0057 \\
	-0.0065 & -0.0057 	& 7.5 \times 10^{-5}
	} , \\ 
	V^e = -A^e &\simeq \pmatr{
	0.99 	& 0.073	& -0.0085 \\
	0.073	& 0.51	& -0.0013 \\
	-0.0085 & -0.0013 & 7.5 \times 10^{-5}
	} , 
\end{split}
\end{equation}

We may immediately compute the branching ratios for all aforementioned meson and lepton decays and neutral meson mass splittings.
The only remaining parameter is the axion scale $ v_{PQ} $, which is only loosely constrained by naturalness arguments to be $ v_{PQ} \sim 10^{12} $ GeV.
In principle, any two measurements of either flavour violation (as discussed in this paper), the axion-photon coupling $ g_{a\gamma} $, or the axion-electron coupling $ g_{ae} $, would be sufficient to overconstrain $ v_{PQ} $ in this model.
Here, $ g_{a\gamma} $ is fixed by $ v_{PQ} $ and the domain wall number $ N_{DW} = 6 $.
In other words, although the charge assignments are very different, the A to Z model will resemble the original DFSZ model in experiments sensitive to $ g_{a\gamma} $, such as haloscopes and helioscopes.
In Table \ref{tab:a2zpredictions} we give the model predictions for some of the most phenomenologically interesting experimental probes. 
We explicitly set $ v_{PQ} = 10^{12} $ GeV when computing the branching ratio.

\begin{table}[!ht]
\centering
\begin{tabular}{crr}
\toprule
	Process & 
	\makecell[r]{Branching ratio \\ ($ v_{PQ} = 10^{12} $ GeV)} &
	Experimental sensitivity \\
\midrule
	$ K^+ \to \pi^+ a $ 	& $2.19 \times 10^{-12}$ & $ \simlt 1 \times 10^{-12} $ (NA62 future) \\
	$ K^0_L \to \pi^0 a $	& $2.29 \times 10^{-12}$ & $ < 5 \times 10^{-8} $ (KOTO) \\
	$ \mu^+ \to e^+ a $ 	& $ 8.3 \times 10^{-13} $ & $ \simlt 5 \times 10^{-9} $ (Mu3e future) \\
\bottomrule
\end{tabular}
\caption{%
	Predictions for axion-induced processes in the A to Z model. 
	Branching ratios are computed assuming $ v_{PQ} = 10^{12} $ GeV, which should be true up to an $ \mathcal{O}(1) $ factor.
}
\label{tab:a2zpredictions}
\end{table}

Predictably, as $ v_{PQ} \sim 10^{12} $ GeV, all processes involving two axion vertices, including meson mixing and $ \mu \to eee $, are heavily suppressed.
For all mesons $ P $, we calculate $ (\Delta m_P)_\mathrm{axion} \sim 10^{-23} - 10^{-24} $ MeV, while for $ \mu \to eee $ the branching ratio is $ \mathcal{O}(10^{-45}) $, essentially undetectable.

In summary, we find that evidence for or against the A to Z model must come primarily from the (non-)observation of $ K^+ \to \pi^+ a $; the NA62 experiment is expected to be able to exclude most of the model's parameter space.
A next-generation experiment could exclude the model definitively.
Secondary sources of interest are decays of $ K^0_L $ and $ \mu^+ $; detecting the A to Z model would require $ v_{PQ} $ to be slightly lower than the natural prediction.
However, two-body decays may be powerful channels for excluding other flavour models, sometimes placing stronger constraints than those from astrophysics, which typically give the strongest limits on $ v_{PQ} $.

\subsection{Decay correlations}

The prominent feature of unified models is correlations between Yukawa couplings of quarks and leptons.
In this A to Z model, $ Y^d \sim Y^e $, up to diagonal Clebsch-Gordan factors. 
Notably, the (2,2) entries differ by a parameter $ x $, which is determined by the fit and acts as a necessary Clebsch-Gordan factor to distinguish the strange quark and muon masses.
Naturally, one expects $ x \sim m_\mu/m_s > 1 $; at the GUT scale, $ m_\mu / m_s \sim 4.5 $.
Now consider the two decays $ K^+ \to \pi^+ a $ and $ \mu^+ \to e^+ a $, which are the most experimentally promising among flavoured axion decays.
Their branching ratios are determined, respectively, by the couplings $ |V^d_{21}|^2 $ and $ |C^e_{21}|^2 = 2 |V^e_{21}|^2 $.
With all other parameters held constant, the dependence on $ x $ of the ratio $ r = |V^e_{21}|^2 / |V^d_{21}|^2$ is well approximated empirically by $ r \approx 6.9 \, e^{-1.8 \sqrt{x}} $.

We then find that the ratio of branching ratios $ R_{\mu/K} $ is given by
\begin{equation}
	R_{\mu/K} \equiv
	\frac{\mathrm{Br}(\mu^+ \to e^+ a)}{\mathrm{Br}(K^+ \to \pi^+ a)}
	\simeq 4.45 \, \frac{|V^e_{21}|^2}{|V^d_{21}|^2} 
	\approx 31 \, e^{-1.8 \sqrt{x}} .
\end{equation}
For the model best fit point $ x = 5.88 $, $ R_{\mu/K} \approx 0.38 $.
Should both of these decays be measured experimentally, such a ratio, which is independent of the axion scale $ v_{PQ} $, is a valuable statistic for constraining the flavour sector of the model, giving immediate information about the high-scale parameters.
For models where $ Y^d \sim Y^e $, typically $ x > 1 $; generically one expects $ R_{\mu/K} < 1 $.
Similar ratios can be considered for other decays of $ K $ or $ B $ mesons and charged leptons.
However, as this requires direct observation of both decays, which are suppressed in both sectors, these are realistically feasible only for more general ALPs.

\section{Conclusion}
\label{sec:conclusion} 

In this paper we have reviewed and extended the phenomenology of flavourful axions, including both standard PQ axions, associated with the solution to the strong $CP$ problem, and also for non-standard axion-like particles (ALPs) which do not care about the strong $CP$ problem but which may generically arise from spontaneously broken symmetries and multiple scalar fields.
We have presented the flavourful axion-fermion and axion-photon couplings both for the standard axion and for ALPs, and shown that they quite naturally are non-diagonal.
Using these couplings, we have calculated the branching ratios for two-body decays of heavy mesons $ K $, $ D $, and $ B $ involving a flavourful axion. 
We have also calculated the mixing between axions and hadronic mesons $K^0$, $D^0$, $ B^0 $ and $B_s^0$ and its consequences, which has not been discussed in the literature before.
These can lead to new contributions to neutral meson mass splitting, meson decays into axions and axion decays into two photons which may be relevant for ALPs. 
We have also discussed charged lepton flavour-violating processes involving final state axions, of the form $\ell_1 \to \ell_2 a (\gamma) $, as well as $ \mu \to eee $ and $ \mu-e $ conversion.

Correlations between observables may arise in specific flavourful axion models.
To illustrate this, we have described the phenomenology of the A to Z Pati-Salam model, which predicts a flavourful QCD axion \cite{Bjorkeroth:2017tsz}, and shown how unification leads to correlations between different flavour-dependent observables, as the down-type quark and charged lepton couplings are very similar. 
Within this model, since the axion arises from the same flavon fields that dictate fermion Yukawa structures, no additional field content is necessary to solve the strong $ CP $ problem, and all axion couplings are fixed by a fit to quark and lepton masses and mixing.

In conclusion, flavourful axions can appear naturally in realistic models and have a rich phenomenology beyond that of the standard KSVZ/DFSZ paradigms. 
In this paper we have attempted to provide the first comprehensive discussion of a number of relevant processes involving flavourful axions, including meson decays and mixing, as well as charged lepton flavour-violating processes. 
For a QCD axion, typically the bounds from such processes are very weak.
However, $K\rightarrow \pi a$ is an ideal channel for looking at these types of decays, especially in specific models such as the A to Z Pati-Salam model, where exactly this type of flavour-violating coupling is the largest.
By comparing multiple flavour-violating processes for both quarks and leptons, one may experimentally probe lepton and quark Yukawa structures which determine their masses and mass ratios. 
Although for QCD axions some of the flavour-violating processes we consider are not competitive, for flavourful ALPs many of them may be important, especially if the symmetry-breaking scale is $10^6$ GeV or less.

\acknowledgments

The authors thank Joerg Jaeckel, Andreas J\"uttner, Enrico Nardi, Giovanni Marco Pruna and Robert Ziegler for helpful discussions and comments.
SFK acknowledges the STFC Consolidated Grant ST/L000296/1 and the European Union's Horizon 2020 Research and Innovation programme under Marie Sk\l{}odowska-Curie grant agreements Elusives ITN No.\ 674896 and InvisiblesPlus RISE No.\ 690575. 
EJC is supported also by InvisiblesPlus RISE No.\ 690575.
FB is supported in part by the INFN ``Iniziativa Specifica'' TAsP-LNF.

\appendix

\section{Axion-meson mixing}
\label{app:sec:mesonmixing}

Kinetic mixing between the axion and neutral mesons (any of the pairs $ K^0-\overbar{K}^0 $, $ D^0-\overbar{D}^0 $, $ B^0-\overbar{B}^0 $) is described by the Lagrangian $ \mathcal{L}_\mathrm{kin}^0 + \mathcal{L}_m^0  $, where
\begin{equation}
\begin{split}
	\mathcal{L}_\mathrm{kin}^0 &= 
		\frac{1}{2} \partial_\mu a \, \partial^\mu a
		+ \frac{1}{2} \partial_\mu P^0 \, \partial^\mu \overbar{P}^0
		- \eta_P \partial_\mu a \, \partial^\mu P^0 
		- \eta_P^\ast \partial_\mu a \, \partial^\mu \overbar{P}^0, \\
	\mathcal{L}_m^0 &= 
		-\frac{1}{2} m_a^2 a^2 
		- m_P^2 P^0 \overbar{P}^0 .
\end{split}
\end{equation}
where $ P^0 $, $ \overbar{P}^0 $ are strong eigenstates.
The superscript $ 0 $ signifies we are not in a diagonal (physical) basis.
We define the $ CP $ eigenstates $ P_1 $ (even) and $ P_2 $ (odd) by
\begin{equation}
	P_1 = \frac{1}{\sqrt{2}} (P^0 + \overbar{P}^0) , \qquad  
	P_2 = \frac{1}{\sqrt{2}} (P^0 - \overbar{P}^0) .
\end{equation}
Inversely, 
\begin{equation}
	P^0 			= \frac{1}{\sqrt{2}} (P_1 + P_2) , \qquad  
	\overbar{P}^0 	= \frac{1}{\sqrt{2}} (P_1 - P_2) .
\end{equation}
In the case of the kaon, the states $ K_{1,2} $ are close (but not exactly equal) to the physical eigenstates $ K_S $ and $ K_L $, so defined by having definite lifetimes in weak decays.
They are given in terms of a small parameter $ \varepsilon_K \sim 10^{-3} $ characterising indirect $ CP $ violation,
\begin{equation}
	K_S = \frac{1}{\sqrt{1 + |\varepsilon_K|^2}} (K_1 + \varepsilon_K K_2) , \qquad  
	K_L = \frac{1}{\sqrt{1 + |\varepsilon_K|^2}} (K_2 + \varepsilon_K K_1) .
\end{equation}
We will neglect such a contribution in this work.
Rewriting $ \mathcal{L}_\mathrm{kin}^0 $ in terms of $ P_{1,2} $, we have
\begin{equation}
\begin{split}
	\mathcal{L}_\mathrm{kin}^0 &= 
		\frac{1}{2} \partial_\mu a \, \partial^\mu a
		+ \frac{1}{2} \partial_\mu P_1 \, \partial^\mu P_1
		- \frac{1}{2} \partial_\mu P_2  \, \partial^\mu P_2
		- \frac{\eta_P + \eta_P^\ast}{\sqrt{2}}  \partial_\mu a \, \partial^\mu P_1 
		- \frac{\eta_P - \eta_P^\ast}{\sqrt{2}}  \partial_\mu a \, \partial^\mu P_2 , \\
		\mathcal{L}_m^0 &= 
		-\frac{1}{2} m_a^2 a^2 
		- m_P^2 P_1^2
		+ m_P^2 P_2^2 .
\end{split}
\end{equation}
Note the wrong sign of the $ P_2 $ diagonal kinetic and mass terms; these can be made canonical by letting $ P_2 \to i P_2 $, which introduces a factor $ i $ in the kinetic mixing term.
This can be absorbed in new couplings $ \eta_{1,2} $, defined by
\begin{equation}
	\eta_1 = \frac{1}{\sqrt{2}} (\eta_P + \eta_P^\ast) ,\qquad
	\eta_2 = - \frac{i}{\sqrt{2}} (\eta_P - \eta_P^\ast) .
\end{equation}
We also define a ``total'' coupling $ \eta^2 \equiv {\eta_1^2 + \eta_2^2} = 2 \eta_P \eta_P^\ast = 2 |\eta_P|^2 $.

We diagonalise the kinetic Lagrangian by transformations
\begin{equation}
	a \to \frac{a}{\sqrt{1-\eta^2}}, \quad
	P_1 \to P_1 + \frac{\eta_1}{\sqrt{1-\eta^2}} a, \quad
	P_2 \to P_2 + \frac{\eta_2}{\sqrt{1-\eta^2}} a. 
\label{app:eq:kineticdiagonalisation}
\end{equation}
The mixing is transferred to the mass matrix, giving
\begin{equation}
	\mathcal{L}_m^0 \to \mathcal{L}_m = - \frac{1}{2} 
	\left[
		m_P^2 (P_1^2 + P_2^2)
		+ \left(\frac{m_a^2 + m_P^2 \eta^2}{1-\eta^2}\right) a^2
		+ \frac{m_P^2 a}{\sqrt{1-\eta^2}} \left( \eta_1 P_1 + \eta_2 P_2 \right)
	\right]
	+ \mathrm{h.c.}.
\label{app:eq:Lmassinduced}
\end{equation}
In matrix form, we may write $ \mathcal{L}_m = -\tfrac{1}{2} \Phi^T M^2_\Phi \Phi $, where $ \Phi = (a, P_1, P_2) $ and 
\begin{equation}
	M^2_\Phi = m_P^2 \pmatr{
		\dfrac{m_a^2}{m_P^2} + \dfrac{\eta^2}{1-\eta^2} & \dfrac{\eta_1}{\sqrt{1-\eta^2}} & \dfrac{\eta_2}{\sqrt{1-\eta^2}} \\
		\dfrac{\eta_1}{\sqrt{1-\eta^2}} & 1 & 0 \\
		\dfrac{\eta_2}{\sqrt{1-\eta^2}} & 0 & 1 
	} .
\label{app:eq:mixingmassmatrix}
\end{equation}
The eigenvalues of $ M^2_\Phi $, corresponding to the physical squared masses, are given to good approximation for small $ \eta $ by
\begin{equation}
	m_a^2 (1 - \eta^2) , \qquad
	m_P^2 (1 + \eta^2) , \qquad
	m_P^2 . 
\end{equation}
Recalling that $ \eta^2 = 2 |\eta_P|^2 $, we conclude that
\begin{equation}
	|\Delta m_P| \equiv |m_{P_1} - m_{P_2}| \simeq m_P \left(\sqrt{1 + 2 |\eta_P|^2} - 1 \right) \simeq |\eta_P|^2 m_P . 
\end{equation}

We have not taken into account a mass difference from SM physics, such as for kaons, where $ K_S $ and $ K_L $ differ by approximately $ 3~\mu $eV.

\section{Heavy meson decay branching ratio}
\label{app:sec:br}

The Feynman rule for the vertex $ (\partial_\mu a) \bar{q}_1 \gamma^\mu q_2 $ defined by the Lagrangian in Eq.~\ref{eq:lagrangian} is 
\begin{equation}
	- i \frac{V^f_{q_1 q_2}}{v_{PQ}} q_\mu \gamma^\mu ,
\end{equation}
where $ q = p_a = p_1 - p_2 $ is the momentum transfer to the axion.
For a two-body decay $ P \to {P^\prime} a $ of a heavy meson $ P = (\bar{q}_P q^\prime) $ into $ {P^\prime} = (\bar{q}_{P^\prime} q^\prime) $, the amplitude may be written
\begin{equation}
	\mathcal{M} = 
		- i\frac{V^f_{q_P q_{P^\prime}}}{v_{PQ}} (p_P - p_{P^\prime})_\mu 
		\bra{{P^\prime}} \bar{q}_P \gamma^\mu q_{{P^\prime}} \ket{P} .
\end{equation}
It depends on a form factor $ f_{+}(q^2) $ encapsulating hadronic physics.
The lightness of the axion means we can safely take the limit $ q^2 \to 0 $, wherein the form factor is defined by the relation
\begin{equation}
	\bra{{P^\prime}} \bar{q}_P \gamma^\mu q_{{P^\prime}} \ket{P} 
		= f_{+}(0) (p_P + p_{P^\prime})^\mu ,
\end{equation}
such that
\begin{equation}
	\mathcal{M} = i \frac{V^f_{q_P q_{P^\prime}}}{v_{PQ}} (m_P^2 - m_{P^\prime}^2) f_{+}(0) .
\end{equation}

The differential decay rate in the rest frame of $ P $ is
\begin{equation}
	\mathrm{d}\Gamma = 
		\frac{1}{32\pi^2} \left| \mathcal{M} \right|^2 
		\frac{|\mathbf{p}_{P^\prime}|}{m_P^2} \mathrm{d}\Omega ,
\end{equation}
with the momentum of decay products $ |\mathbf{p}_{P^\prime}| = |\mathbf{p}_a| $ given by
\begin{equation}
	|\mathbf{p}_{P^\prime}| 
	= |\mathbf{p}_a| 
	= \frac{\left[ (m_P^2 - (m_{P^\prime} + m_a)^2) (m_P^2 - (m_{P^\prime} - m_a)^2) \right]^{1/2}}{2m_P} 
	\stackrel{(m_a \ll m_{P^\prime})}{\approx}
	\frac{m_P^2 - m_{P^\prime}^2}{2m_P} ,
\end{equation}
Integrating over the solid angle $ \Omega $ yields a factor $ 4 \pi $, arriving at
\begin{equation}
	\Gamma(P \to {P^\prime} a) = 
		\frac{1}{16\pi} \frac{\big|V^f_{q_P q_{P^\prime}}\big|^2}{v_{PQ}^2}
		m_P^3 \left( 1 - \frac{m_{P^\prime}^2}{m_P^2} \right)^3
		\left| f_{+}(0) \right|^2.
\end{equation}

\section{Couplings in the A to Z Pati-Salam Model}
\label{app:sec:couplings}

\subsubsection*{Superpotential}

The effective Yukawa superpotential below the GUT scale, once messengers $ X $ have been integrated out, is given by
\begin{equation}
\begin{split}
	W_Y^\mathrm{eff} &= 
	\lambda_3 (F \cdot h_3) F^c_3
	+ \lambda_{1u} \frac{(F \cdot \phi_1^u) h_u F^c_1}{\braket{\Sigma_u}}
	+ \lambda_{2u} \frac{(F \cdot \phi_2^u) h_u F^c_2}{\braket{\Sigma_u}}
	\\ & \qquad
	+ \lambda_{1d} \frac{(F \cdot \phi_1^d) h_d F^c_1}{\braket{\Sigma^d_{15}}}
	+ \lambda_{2d} \frac{(F \cdot \phi_2^d) h^d_{15} F^c_2}{\braket{\Sigma_d}}
	+ \lambda_{ud} \frac{(F \cdot \phi_1^u) h_d F^c_1}{\braket{\Sigma_d}} ,
\end{split}
\end{equation}
with explicit couplings $ \lambda $, which are naturally $ \mathcal{O}(1) $ and assumed real by a $ CP $ symmetry at high scale.
In the corresponding Lagrangian, the fermion part of the chiral superfields $ F $, $ F^c_i $ are denoted $ f $, $ f^c_i $, respectively.%
\footnote{
	To be precise: $ f, f^c_i $ are Weyl fermions, by definition transforming as left-handed fields.
	In other words, $ f^c_i $ are the left-handed components of a weak $ SU(2)_L $ singlet. 
}
These are the familiar SM fermions as well as a set of right-handed neutrinos.
The light Higgs scalar doublets keep the same notation as their corresponding superfield.%
\footnote{
	This is rather imprecise but tolerable, as the Higgs sector is not relevant to the PQ mechanism, and fields are anyway replaced by their VEVs eventually.
}
The fields $ \Sigma $ acquire high-scale VEVs which give dynamical masses to the $ X $ messengers in the renormalisable theory, expected to be $ \mathcal{O}(M_\mathrm{GUT}) $.

\subsubsection*{Goldstone field}

The central actors in the flavoured axion model are the $ A_4 $ triplet flavons $ \phi $.
Taking only the scalar part of superfields $ \phi $, we let 
\begin{equation}
	\phi_i \to \varphi_i = \tfrac{1}{\sqrt{2}} \left(\braket{\varphi} + \rho_\varphi \right)_i e^{i a_\varphi / v_\varphi} , 
	\qquad
	\braket{\varphi} \equiv v_\varphi \mathbf{x}_\varphi ,
\end{equation}
where we have expanded around the flavon VEVs, noting that each $ \braket{\varphi} $ consists of a scale $ v $ and direction $ \mathbf{x} $ in $ A_4 $ space.
The VEVs are aligned according to the CSD(4) prescription, i.e.
\begin{equation}
\begin{aligned}
	\mathbf{x}_{\varphi_1^u} &= (0,1,1), &\qquad
	\mathbf{x}_{\varphi_1^u} &= (1,4,2), \\
	\mathbf{x}_{\varphi_1^u} &= (1,0,0), &\qquad
	\mathbf{x}_{\varphi_1^u} &= (0,1,0).
\end{aligned}
\end{equation}
The radial fields $ \rho_\varphi $ are very heavy and phenomenologically uninteresting, so will be neglected henceforth.
The phase fields $ a_\varphi $ are not independent, but related by the single $ U(1) $ rephasing symmetry.
We identify the Goldstone (or axion) field $ a $ by
\begin{equation}
	a \equiv \sum_\varphi x_\varphi \frac{v_\varphi a_\varphi}{v_{PQ}} ,
	\qquad 
	v_{PQ}^2 \equiv \sum_\varphi x_\varphi^2 v_\varphi^2 .
\end{equation}
Component fields are given by
\begin{equation}
	a_\varphi = \frac{x_\varphi v_\varphi}{v_{PQ}} a .
\end{equation}

\subsubsection*{Lagrangian (SUSY basis)}

The Yukawa Lagrangian may thus be written as
\begin{equation}
\begin{split}
	- \mathcal{L}_Y &\supset 
	\lambda_3 f h_3 f^c_3 
	+ \frac{\lambda_{1u}}{\sqrt{2} \braket{\Sigma_u}} (f \cdot \braket{\varphi_1^u}) h_u f^c_1 \Exp{\frac{i x_{\varphi_1^u}  a}{v_{PQ}}} 
	\\ &\qquad 
	+ \frac{\lambda_{2u}}{\sqrt{2} \braket{\Sigma_u}} (f \cdot \braket{\varphi_2^u}) h_u f^c_2 \Exp{\frac{i x_{\varphi_2^u}  a}{v_{PQ}}} 
	+ \frac{\lambda_{1d}}{\sqrt{2} \braket{\Sigma_{15}^d}} (f \cdot \braket{\varphi_1^d}) h_d f^c_1 \Exp{\frac{i x_{\varphi_1^d}  a}{v_{PQ}}} 
	\\&\qquad 
	+ \frac{\lambda_{2d}}{\sqrt{2} \braket{\Sigma_d}} (f \cdot \braket{\varphi_2^d}) h_{15}^d f^c_2 \Exp{\frac{i x_{\varphi_2^d}  a}{v_{PQ}}} 
	+ \frac{\lambda_{ud}}{\sqrt{2} \braket{\Sigma_d}} (f \cdot \braket{\varphi_1^u}) h_d f^c_1 \Exp{\frac{i x_{\varphi_1^u}  a}{v_{PQ}}} 
	\\&\qquad
	+ \mathcal{O}(\rho_\varphi)
	+ \, \mathrm{h.c.}.
\end{split}
\label{app:eq:LY0}
\end{equation}
Let us make the SM field components of the PS fields $ f, f^c_i $ explicit: $ f \to (Q, L) $, $ f^c_i \to (u^c_i, d^c_i) $.
Below the EWSB scale, $ Q $ and $ L $ further decompose into $ (u_L, d_L) $ and $ (\nu_L, e_L) $, respectively.
In addition, $ h_u \to v_u $, $ h_d, h_{15}^d \to v_d $, with some small mixing assumed between Higgs bi-doublets to give the MSSM 2HDM;
we assume the effects of this mixing are negligible.
The fields $ \Sigma $ acquire real VEVs, with magnitudes generically written $ v_\Sigma $, i.e.
\begin{equation}
	\braket{\Sigma_u} \to v_{\Sigma_u}, \quad
	\braket{\Sigma_{15}^d} \to v_{\Sigma_d}, \quad
	\braket{\Sigma_d} \to v_{\Sigma_d} ,
\end{equation}
The interplay between the singlet $ \Sigma_d $ and adjoint $ \Sigma_{15}^d $ also provides Clebsch-Gordan factors which are different for quarks and leptons.
To account for the split between down-type quarks and charged leptons, we reparametrise the couplings $ \lambda $ in the charged lepton sector, so
$ \lambda_{1d} \to \tilde{\lambda}_{1d} $,
$ \lambda_{2d} \to \tilde{\lambda}_{2d} $, and
$ \lambda_{ud} \to \tilde{\lambda}_{ud} $.

\subsubsection*{Lagrangian (left-right basis)}

It is also convenient to work in the left-right (LR) basis, in terms of Weyl fermions $ u_{L,R} $, $ d_{L,R} $, $ e_{L,R} $, and $ \nu_{L,R} $.
This amounts to nothing more than taking the Hermitian conjugate of the terms in Eq.~\ref{app:eq:LY0}.
With all above considerations taken into account, the Lagrangian becomes
\begin{equation}
\begin{split}
	- \mathcal{L}_{Y} &=
	\lambda_3 (\bar{f} \cdot \braket{h_3}^\ast) f_{R3} 
	+ \frac{\lambda_{1u} v_u}{\sqrt{2} v_{\Sigma_u}} (\bar{u}_L \cdot \braket{\varphi_1^u}^\ast) u_{R1} \Exp{\frac{-i x_{\varphi_1^u} a}{v_{PQ}}} 
	\\ &\qquad 
	+ \frac{\lambda_{2u} v_u}{\sqrt{2} v_{\Sigma_u}} (\bar{u}_L \cdot \braket{\varphi_2^u}^\ast) u_{R2} \Exp{\frac{-i x_{\varphi_2^u} a}{v_{PQ}}} 
	+ \frac{\lambda_{1d} v_d}{\sqrt{2} v_{\Sigma_d}} (\bar{d}_L \cdot \braket{\varphi_1^d}^\ast) d_{R1} \Exp{\frac{-i x_{\varphi_1^d} a}{v_{PQ}}} 
	\\&\qquad 
	+ \frac{\lambda_{2d} v_d}{\sqrt{2} v_{\Sigma_d}} (\bar{d}_L \cdot \braket{\varphi_2^d}^\ast) d_{R2} \Exp{\frac{-i x_{\varphi_2^d}  a}{v_{PQ}}} 
	+ \frac{\lambda_{ud} v_d}{\sqrt{2} v_{\Sigma_d}} (\bar{d}_L \cdot \braket{\varphi_1^u}^\ast) d_{R1} \Exp{\frac{-i x_{\varphi_1^u}  a}{v_{PQ}}} 
	\\&\qquad
	+ \left\{
		d_L \to e_L, 
		d_R \to e_R, 
		\lambda_{1d} \to \tilde{\lambda}_{1d}, 
		\lambda_{2d} \to \tilde{\lambda}_{2d}, 
		\lambda_{ud} \to \tilde{\lambda}_{ud}
	\right\} 
	+ \, \mathrm{h.c.}.
\end{split}
\end{equation}
This rather hefty expression can be put in a more conventional format by 1) expanding the $ A_4 $ triplet products like $ \overbar{Q}\cdot \braket{\varphi} $, such that we may write the couplings as matrices, and 2) noting that each term must be PQ-invariant, allowing us to replace the flavon PQ charges with those of the SM fermions.%
\footnote{
	Note that under hermitian conjugation, the PQ charges change sign, i.e $ x_{f^c} \equiv - x_{f_R} $.
	For consistency, we will always specify which Weyl fermion we are referring to.
}
Moreover, all $ \lambda $ are real by an assumed $ CP $ symmetry at high scales.

\subsubsection*{Lagrangian (condensed linear basis)}

Collecting all free parameters, we have
\begin{equation}
\begin{split}
	- \mathcal{L}_Y &=
	e^{i \frac{a}{v_{PQ}} (x_{u_{Li}} - x_{u_{Rj}})} M^u_{ij} \, \bar{u}_{Li} u_{Rj} 
	+ e^{i \frac{a}{v_{PQ}} (x_{d_{Li}} - x_{d_{Rj}})} M^d_{ij} \, \bar{d}_{Li} d_{Rj} 
	\\ &\qquad 
	+ e^{i \frac{a}{v_{PQ}} (x_{e_{Li}} - x_{e_{Rj}})} M^e_{ij} \, \bar{e}_{Li} e_{Rj} 
	+ \mathrm{h.c.}.
\end{split}
\end{equation}
The coupling matrices are given exactly by
\begin{equation}
\begin{split}
	M^u_{ij} &= \frac{v_u}{\sqrt{2} v_{\Sigma_u}} \left( 
		\lambda_{ju}^\ast \braket{\varphi_j^u}_i^\ast
		+ \delta_{j3} \lambda_3^\ast \Theta_i(u) 
		\right) ,
	\\	
	M^d_{ij} &= \frac{v_d}{\sqrt{2} v_{\Sigma_d}} \left( 
		\lambda_{jd}^\ast \braket{\varphi_j^d}_i^\ast 
		+ \delta_{j1} \lambda_{ud}^\ast \braket{\varphi_1^u}_i^\ast 
		+ \delta_{j3} \lambda_3^\ast \Theta_i(d) 
		\right) ,
	\\
	M^e_{ij} &= \frac{v_d}{\sqrt{2} v_{\Sigma_d}} \left( 
		\tilde{\lambda}_{jd}^\ast \braket{\varphi_j^d}_i^\ast 
		+ \delta_{j1} \tilde{\lambda}_{ud}^\ast \braket{\varphi_1^u}_i^\ast 
		+ \delta_{j3} \lambda_3^\ast \Theta_i(d) 
		\right) ,
\end{split}
\end{equation}
where $ \Theta_i(f) $ is a function taking into account the VEV alignment of the $ A_4 $ triplet $ h_3 $, as well as mixing effects between various Higgs doublets.
It traces its origin to the term $ \bar{f} h_3 f_{R3} $, and fixes the third column of the Yukawa matrices.
As $ F^c_3 $ is uncharged under $ U(1)_{PQ} $, the exact form of $ \Theta(h_3,f) $ has only marginal relevance for axion physics.
We refer the interested reader to the original ``A to Z'' paper \cite{King:2014iia} 
for a fuller discussion on Higgs mixing and the origin of the third family couplings.

The above expressions, while precise, are not very illustrative.
In explicit matrix form, we have
\begin{equation}
\begin{split}
	M^u = v_u \pmatr{
		0 & b & \epsilon_{13} c \\
		a  & 4 b & \epsilon_{23} c \\
		a  & 2 b & c \\
	}, 
	~~
	M^d = v_d \pmatr{
		y_d^0  & 0 & 0 \\
		B y_d^0  & y_s^0  & 0 \\
		B y_d^0  & 0 & y_b^0
	}, 
	~~
	M^e = v_d \pmatr{
		-(y_d^0/3)  & 0 & 0 \\
		B y_d^0  & x y_s^0  & 0 \\
		B y_d^0  & 0 & y_b^0
	} ,
\end{split}
\end{equation}
with dimensionless parameters defined by
\begin{equation}
\begin{aligned}
	a &= \frac{\lambda_{1u} v_{\varphi_1^u}}{\sqrt{2} v_{\Sigma_u}}, &
	b &= \frac{\lambda_{2u} v_{\varphi_2^u}}{\sqrt{2} v_{\Sigma_u}}, &
	c &= \lambda_3 \Theta_3(u), & 
	\epsilon_{i3} &= \frac{\Theta_i(u)}{\Theta_3(u)}
	\\
	y_d^0 &= \frac{\lambda_{1d} v_{\varphi_1^d}}{\sqrt{2} v_{\Sigma_d}}, &
	y_s^0 &= \frac{\lambda_{1d} v_{\varphi_2^d}}{\sqrt{2} v_{\Sigma_d}}, &
	y_b^0 &= \lambda_3 \Theta_3(d) , &
	B &= \frac{\lambda_{ud} v_{\varphi_1^u}}{\lambda_{1d} v_{\varphi_1^d}} . \\
\end{aligned}
\end{equation}

\subsubsection*{Lagrangian (derivative basis)}

We perform an axion-dependent rotation of the fermion fields to replace the linear couplings with derivative ones; the anomaly term is also induced.
Extending the Lagrangian to include the fermion kinetic terms,
$ \sum_f (\bar{f}_{Li} \slashed{\partial} f_{Li} + \bar{f}_{Ri} \slashed{\partial} f_{Ri}) $,
we let
\begin{equation}
	f_{Li} \to e^{i \frac{a}{v_{PQ}} x_{f_{Li}}} f_{Li} , \qquad
	f_{Ri} \to e^{i \frac{a}{v_{PQ}} x_{f_{Ri}}} f_{Ri} ,
\end{equation}
resulting in
\begin{equation}
\begin{split}
	\mathcal{L} &= 
	i \sum_{f = u, d, e} 
	\left(
		\bar{f}_{Li} \slashed{\partial} f_{Li}
		+ \bar{f}_{Ri} \slashed{\partial} f_{Ri} 
	\right)
	- \frac{\partial_\mu a}{v_{PQ}} 
	\sum_{f = u, d, e} 
	\left[
		x_{f_{Li}} \bar{f}_{Li} \gamma^\mu f_{Li}
		+ x_{f_{Ri}} \bar{f}_{Ri} \gamma^\mu f_{Ri}
	\right]
	\\ &\qquad 
	- M^u_{ij} \, \bar{u}_{Li} u_{Rj} 
	- M^d_{ij} \, \bar{d}_{Li} d_{Rj} 
	- M^e_{ij} \, \bar{e}_{Li} e_{Rj}
	+ \mathrm{h.c.}
	+ \mathrm{anomaly}.
\end{split}
\end{equation}

We rotate to the mass basis by unitary transformations
$ u_L \to U_Q u_L $,
$ d_L \to U_Q d_L $,
$ e_L \to U_L e_L $,
$ f_R \to V_f f_R $,
such that the mass terms become
\begin{equation}
\begin{split}
	\mathcal{L}_m &=
	M^u_{ij} \, \bar{u}_{Li} u_{Rj} 
	+ M^d_{ij} \, \bar{d}_{Li} d_{Rj} 
	+ M^e_{ij} \, \bar{e}_{Li} e_{Rj}
	+ \mathrm{h.c.}
	\\ &\qquad \to
	m^u_i \delta_{ij} \bar{u}_{Li} u_{Rj} 
	+ (V_\mathrm{CKM})_{ik} m^d_k \delta_{kj} \bar{d}_{Li} d_{Rj} 
	+ m^e_i \delta_{ij} \bar{e}_{Li} e_{Rj} 
	+ \mathrm{h.c.},
\end{split}
\end{equation}
where by definition $ m^f \equiv U_f^\dagger M^f V_f $, $ U_Q \equiv U_u $, and $ V_\mathrm{CKM} \equiv U_u^\dagger U_d $.

\subsubsection*{Derivative couplings}

The axion-fermion derivative couplings become
\begin{equation}
\begin{split}
	\mathcal{L}_\partial = 
	- \frac{\partial_\mu a}{v_{PQ}} 
	\sum_{f = u, d, e} 
	\left[
		\bar{f}_L (U_f^\dagger x_{f_L} U_f) \gamma^\mu f_L
		+ \bar{f}_R (V_f^\dagger x_{f_R} V_f) \gamma^\mu f_R
	\right] 
	+ \mathrm{h.c.},
\end{split}
\end{equation}
where now $ f_L $, $ f_R $ are vectors and $ x_{f_L} $, $ x_{f_R} $ are diagonal $ 3\times3 $ matrices. 
We define the coupling matrices
$ X_L \equiv U_f^\dagger x_{f_L} U_f$ and $ X_R \equiv V_f^\dagger x_{f_R} V_f $,
and note that, since charges $ x_{f} $ are real, $ X_L = X_L^\dagger $ and $ X_R = X_R^\dagger $.
In terms of Dirac spinors, 
\begin{equation}
	\mathcal{L}_\partial = 
	- \frac{\partial_\mu a}{v_{PQ}} 
	\sum_{f = u, d, e} 
	\bar{f} \gamma^\mu (V_f - A_f \gamma_5) f,
\end{equation}
where
\begin{equation}
\begin{split}
	V_f &= \frac{1}{2} (X_L + X_R) = \frac{1}{2} \left( U_f^\dagger x_{f_L} U_f + V_f^\dagger x_{f_R} V_f \right) , \\
	A_f &= \frac{1}{2} (X_L - X_R) = \frac{1}{2} \left( U_f^\dagger x_{f_L} U_f - V_f^\dagger x_{f_R} V_f \right) .
\end{split}
\end{equation}

\section{Couplings in A to Z: numerical fit}
\label{app:sec:numericalfit}

The best fit parameters, as well as a Bayesian 95\% credible interval, are given in Tables~\ref{app:tab:outputleptons} (leptons) and \ref{app:tab:outputquarks} (quarks).
The corresponding best fit input parameters are given in Table~\ref{app:tab:parameters}.
We fit the model to data at the GUT scale.
The running from low to high scale was performed, assuming the MSSM, in \cite{Antusch:2013jca}.
They parametrise threshold corrections by a series of dimensionless parameters $ \eta_i $.
All but one ($ \bar{\eta}_b $) were set to zero, and choosing $ \bar{\eta}_b = -0.24 $ to account for the small GUT-scale difference between $ b $ and $ \tau $ masses. 

\begin{table}[htb]
\centering
\begin{tabular}{ l c D{@}{\ \to \ }{6,6} c@{\hskip 3pt} c D{@}{\ \to \ }{6,6} }
\toprule
	\multirow{2}{*}{Observable}& \multicolumn{2}{c}{Data} && \multicolumn{2}{c}{Model} \\
\cmidrule{2-3} \cmidrule{5-6}
	& Central value & \multicolumn{1}{c}{1$\sigma$ range}  && Best fit & \multicolumn{1}{c}{Interval} \\
\midrule
	$\theta_{12}^\ell \, /^\circ$ 	& 33.57 & 32.81 @ 34.32 && 32.88 & 32.72 @ 34.23 \\ 
	$\theta_{13}^\ell \, /^\circ$ 	& 8.460 & 8.310 @ 8.610 && 8.611 & 8.326 @ 8.882 \\  
	$\theta_{23}^\ell \, /^\circ$ 	& 41.75 & 40.40 @ 43.10 && 39.27 & 37.35 @ 40.11 \\ 
	$\delta^\ell \, /^\circ$ 		& 261.0 & 202.0 @ 312.0 && 242.6 & 231.4 @ 249.9 \\
	$y_e$  $/ 10^{-5}$ 				& 1.004 & 0.998 @ 1.010 && 1.006 & 0.911 @ 1.015 \\ 
	$y_\mu$  $/ 10^{-3}$ 			& 2.119 & 2.106 @ 2.132 && 2.116 & 2.093 @ 2.144 \\ 
	$y_\tau$  $/ 10^{-2}$	 		& 3.606 & 3.588 @ 3.625 && 3.607 & 3.569 @ 3.643 \\ 
	$\Delta m_{21}^2 \, / 10^{-5} \, \mathrm{eV}^2 $ 
									& 7.510  & 7.330 @ 7.690 && 7.413 & 7.049 @ 7.762 \\
	$\Delta m_{31}^2 \, / 10^{-3} \, \mathrm{eV}^2 $ 
									& 2.524 & 2.484 @ 2.564 && 2.540 & 2.459 @ 2.616 \\
	$m_1$ /meV 						& & \multicolumn{1}{c}{} && 0.187 & 0.022 @ 0.234 \\ 
	$m_2$ /meV						& & \multicolumn{1}{c}{} && 8.612 & 8.400 @ 8.815 \\ 
	$m_3$ /meV 						& & \multicolumn{1}{c}{} && 50.40 & 49.59 @ 51.14 \\
	$\sum m_i$ /meV 				& & \multicolumn{1}{c}{$<$ 230 \cite{Ade:2015xua}} && 59.20 & 58.82 @ 60.19 \\
	$ \alpha_{21} $ 				& & \multicolumn{1}{c}{} && 10.4 & -38.0 @ 70.1 \\
	$ \alpha_{31} $ 				& & \multicolumn{1}{c}{} && 272.1 & 218.2 @ 334.0 \\
	$ m_{\beta\beta} $ /meV 		& & \multicolumn{1}{c}{} && 1.940 & 1.892 @ 1.998 \\
\bottomrule
\end{tabular}
\caption{%
	Model predictions in the lepton sector, at the GUT scale.
	We set $\tan \beta = 5$, $ M_\mathrm{SUSY} = 1 $ TeV and $ \bar{\eta}_b = -0.24 $.
	The model interval is a Bayesian 95\% credible interval. 
}
\label{app:tab:outputleptons}
\end{table}

\begin{table}[htb]
\centering
\begin{tabular}{ l c D{@}{\ \to \ }{6,6} c@{\hskip 3pt} c D{@}{\ \to \ }{6,6} }
\toprule
	\multirow{2}{*}{Observable} & \multicolumn{2}{c}{Data} && \multicolumn{2}{c}{Model} \\
\cmidrule{2-3} \cmidrule{5-6}
	& Central value & \multicolumn{1}{c}{$1\sigma$ range} && Best fit & \multicolumn{1}{c}{Interval} \\
\midrule
	$\theta_{12}^q \, /^\circ$ & 13.03 & 12.99 @ 13.07 && 13.04 & 12.94 @ 13.11 \\	
	$\theta_{13}^q \, /^\circ$ & 0.1471 & 0.1418 @ 0.1524 && 0.1463 & 0.1368 @ 0.1577 \\
	$\theta_{23}^q \, /^\circ$ & 1.700 & 1.673 @ 1.727 && 1.689 & 1.645 @ 1.753 \\	
	$\delta^q \, /^\circ$ & 69.22 & 66.12 @ 72.31 && 68.85 & 63.00 @ 75.24 \\
	$y_u \, / 10^{-6}$ & 2.982 & 2.057 @ 3.906 && 3.038 & 1.098 @ 4.957  \\	
	$y_c \, / 10^{-3}$ & 1.459 & 1.408 @ 1.510 && 1.432 & 1.354 @ 1.560 \\	
	$y_t$  			   & 0.544 & 0.537 @ 0.551 && 0.545 & 0.530 @ 0.558 \\
	$y_d \, / 10^{-5}$ & 2.453 & 2.183 @ 2.722 && 2.296 & 2.181 @ 2.966 \\	
	$y_s \, / 10^{-4}$ & 4.856 & 4.594 @ 5.118 && 4.733 & 4.273 @ 5.379 \\
	$y_b$   		   & 3.616 & 3.500 @ 3.731 && 3.607 & 3.569 @ 3.643 \\
\bottomrule
\end{tabular}
\caption{%
	Model predictions in the quark sector at the GUT scale.
	We set $\tan\beta = 5$, $M_\mathrm{SUSY} = 1$ TeV and $\bar{\eta}_b = -0.24$. 
	The model interval is a Bayesian 95\% credible interval.
}
\label{app:tab:outputquarks}
\end{table}

\begin{table}[ht]
\centering
\renewcommand{\arraystretch}{1.1}
\begin{tabular}[t]{lr}
\toprule
	Parameter & Value \\ 
\midrule
	$a \, /10^{-5}$ & $1.246 \, e^{4.047i}$ \\
	$b \, /10^{-3}$ & $3.438 \, e^{2.080i}$ \\
	$c$ 			& $-0.545$ \\
	$y_d^0 \, /10^{-5}$ & $3.053 \, e^{4.816i}$ \\
	$y_s^0 \, /10^{-4}$ & $3.560 \, e^{2.097i}$ \\
	$y_b^0 \, /10^{-2}$ & $3.607$ \\
\bottomrule
\end{tabular}
\hspace*{0.3cm}
\begin{tabular}[t]{lr}
\toprule
	Parameter & Value \\ 
\midrule
	$\epsilon_{13} \, /10^{-3}$ & $6.215 \, e^{2.434i}$ \\
	$\epsilon_{23} \, /10^{-2}$ & $2.888 \, e^{3.867i}$ \\
	$B$ & $10.20 \, e^{2.777i}$ \\ 
	$x$	& $5.880$ \\
\bottomrule
\end{tabular}
\hspace*{0.3cm}
\begin{tabular}[t]{lr}
\toprule
	Parameter & Value \\ 
\midrule
	$m_a$ /meV & $3.646$ \\
	$m_b$ /meV & $1.935$ \\
	$m_c$ /meV & $1.151$ \\
	$\eta$ & $2.592$ \\
	$\xi$ & $2.039$ \\
\bottomrule
\end{tabular}
\caption{Best fit input parameter values.} 
\label{app:tab:parameters}
\end{table}


\clearpage

\begin{thebibliography}{99}
\bibitem{Peccei:1977hh}
  R.~D.~Peccei and H.~R.~Quinn,
  Phys.\ Rev.\ Lett.\  {\bf 38} (1977) 1440.


\bibitem{Wilczek:1977pj}
  F.~Wilczek,
  Phys.\ Rev.\ Lett.\  {\bf 40} (1978) 279.


\bibitem{Weinberg:1977ma}
  S.~Weinberg,
  Phys.\ Rev.\ Lett.\  {\bf 40} (1978) 223.


\bibitem{Kim:1979if}
  J.~E.~Kim,
  Phys.\ Rev.\ Lett.\  {\bf 43} (1979) 103.


\bibitem{Shifman:1979if}
  M.~A.~Shifman, A.~I.~Vainshtein and V.~I.~Zakharov,
  Nucl.\ Phys.\ B {\bf 166} (1980) 493.


\bibitem{Dine:1981rt}
  M.~Dine, W.~Fischler and M.~Srednicki,
  Phys.\ Lett.\  {\bf 104B} (1981) 199.


\bibitem{Zhitnitsky:1980tq}
  A.~R.~Zhitnitsky,
  Sov.\ J.\ Nucl.\ Phys.\  {\bf 31} (1980) 260
   [Yad.\ Fiz.\  {\bf 31} (1980) 497].


\bibitem{Preskill:1982cy}
  J.~Preskill, M.~B.~Wise and F.~Wilczek,
  Phys.\ Lett.\ B {\bf 120} (1983) 127
   [Phys.\ Lett.\  {\bf 120B} (1983) 127].


\bibitem{Abbott:1982af}
  L.~F.~Abbott and P.~Sikivie,
  Phys.\ Lett.\ B {\bf 120} (1983) 133
   [Phys.\ Lett.\  {\bf 120B} (1983) 133].


\bibitem{Dine:1982ah}
  M.~Dine and W.~Fischler,
  Phys.\ Lett.\ B {\bf 120} (1983) 137
   [Phys.\ Lett.\  {\bf 120B} (1983) 137].


\bibitem{Kim:2008hd}
  J.~E.~Kim and G.~Carosi,
  Rev.\ Mod.\ Phys.\  {\bf 82} (2010) 557
  [arXiv:0807.3125 [hep-ph]].


\bibitem{Holman:1992us}
  R.~Holman, S.~D.~H.~Hsu, T.~W.~Kephart, E.~W.~Kolb, R.~Watkins and L.~M.~Widrow,
  Phys.\ Lett.\ B {\bf 282} (1992) 132
  [hep-ph/9203206].


\bibitem{Kamionkowski:1992mf}
  M.~Kamionkowski and J.~March-Russell,
  Phys.\ Lett.\ B {\bf 282} (1992) 137
  [hep-th/9202003].


\bibitem{Barr:1992qq}
  S.~M.~Barr and D.~Seckel,
  Phys.\ Rev.\ D {\bf 46} (1992) 539.


\bibitem{Chun:1992bn}
  E.~J.~Chun and A.~Lukas,
  Phys.\ Lett.\ B {\bf 297} (1992) 298
  [hep-ph/9209208].


\bibitem{BasteroGil:1997vn}
  M.~Bastero-Gil and S.~F.~King,
  Phys.\ Lett.\ B {\bf 423} (1998) 27
  [hep-ph/9709502].


\bibitem{Babu:2002ic}
  K.~S.~Babu, I.~Gogoladze and K.~Wang,
  Phys.\ Lett.\ B {\bf 560} (2003) 214
  [hep-ph/0212339].


\bibitem{Dias:2002hz}
  A.~G.~Dias, V.~Pleitez and M.~D.~Tonasse,
  Phys.\ Rev.\ D {\bf 69} (2004) 015007
  [hep-ph/0210172].


\bibitem{Dias:2002gg}
  A.~G.~Dias, V.~Pleitez and M.~D.~Tonasse,
  Phys.\ Rev.\ D {\bf 67} (2003) 095008
  [hep-ph/0211107].


\bibitem{Dias:2007vx}
  A.~G.~Dias, E.~T.~Franco and V.~Pleitez,
  Phys.\ Rev.\ D {\bf 76} (2007) 115010
  [arXiv:0708.1009 [hep-ph]].


\bibitem{Harigaya:2013vja}
  K.~Harigaya, M.~Ibe, K.~Schmitz and T.~T.~Yanagida,
  Phys.\ Rev.\ D {\bf 88} (2013) no.7,  075022
  [arXiv:1308.1227 [hep-ph]].


\bibitem{Cheung:2010hk}
  C.~Cheung,
  JHEP {\bf 1006} (2010) 074
  [arXiv:1003.0941 [hep-ph]].


\bibitem{DiLuzio:2017tjx}
  L.~Di Luzio, E.~Nardi and L.~Ubaldi,
  Phys.\ Rev.\ Lett.\  {\bf 119} (2017) no.1,  011801
  [arXiv:1704.01122 [hep-ph]].


\bibitem{Wilczek:1982rv}
  F.~Wilczek,
  Phys.\ Rev.\ Lett.\  {\bf 49} (1982) 1549.


\bibitem{Babu:1992cu}
  K.~S.~Babu and S.~M.~Barr,
  Phys.\ Lett.\ B {\bf 300} (1993) 367
  [hep-ph/9212219].


\bibitem{Albrecht:2010xh}
  M.~E.~Albrecht, T.~Feldmann and T.~Mannel,
  JHEP {\bf 1010} (2010) 089
  [arXiv:1002.4798 [hep-ph]].


\bibitem{Celis:2014iua}
  A.~Celis, J.~Fuentes-Martin and H.~Serodio,
  Phys.\ Lett.\ B {\bf 741} (2015) 117
  [arXiv:1410.6217 [hep-ph]].


\bibitem{Ahn:2014gva}
  Y.~H.~Ahn,
  Phys.\ Rev.\ D {\bf 91} (2015) 056005
  [arXiv:1410.1634 [hep-ph]].


\bibitem{Ema:2016ops}
  Y.~Ema, K.~Hamaguchi, T.~Moroi and K.~Nakayama,
  JHEP {\bf 1701} (2017) 096
  [arXiv:1612.05492 [hep-ph]].


\bibitem{Calibbi:2016hwq}
  L.~Calibbi, F.~Goertz, D.~Redigolo, R.~Ziegler and J.~Zupan,
  Phys.\ Rev.\ D {\bf 95} (2017) no.9,  095009
  [arXiv:1612.08040 [hep-ph]].


\bibitem{Choi:2017gpf}
  K.~Choi, S.~H.~Im, C.~B.~Park and S.~Yun,
  JHEP {\bf 1711} (2017) 070
  [arXiv:1708.00021 [hep-ph]].


\bibitem{Arias-Aragon:2017eww}
  F.~Arias-Aragon and L.~Merlo,
  JHEP {\bf 1710} (2017) 168
  [arXiv:1709.07039 [hep-ph]].


\bibitem{Linster:2018avp}
  M.~Linster and R.~Ziegler,
  arXiv:1805.07341 [hep-ph].


\bibitem{King:2013eh}
  S.~F.~King and C.~Luhn,
  Rept.\ Prog.\ Phys.\  {\bf 76} (2013) 056201
  [arXiv:1301.1340 [hep-ph]].


\bibitem{King:2015aea}
  S.~F.~King,
  J.\ Phys.\ G {\bf 42} (2015) 123001
  [arXiv:1510.02091 [hep-ph]].


\bibitem{King:2017guk}
  S.~F.~King,
  Prog.\ Part.\ Nucl.\ Phys.\  {\bf 94} (2017) 217
  [arXiv:1701.04413 [hep-ph]].


\bibitem{King:2014nza}
  S.~F.~King, A.~Merle, S.~Morisi, Y.~Shimizu and M.~Tanimoto,
  New J.\ Phys.\  {\bf 16} (2014) 045018
  [arXiv:1402.4271 [hep-ph]].


\bibitem{Bjorkeroth:2017tsz}
  F.~Björkeroth, E.~J.~Chun and S.~F.~King,
  Phys.\ Lett.\ B {\bf 777} (2018) 428
  [arXiv:1711.05741 [hep-ph]].


\bibitem{King:2014iia}
  S.~F.~King,
  JHEP {\bf 1408} (2014) 130
  [arXiv:1406.7005 [hep-ph]].


\bibitem{Bjorkeroth:2015uou}
  F.~Björkeroth, F.~J.~de Anda, I.~de Medeiros Varzielas and S.~F.~King,
  Phys.\ Rev.\ D {\bf 94} (2016) no.1,  016006
  [arXiv:1512.00850 [hep-ph]].


\bibitem{Nelson:1983zb}
  A.~E.~Nelson,
  Phys.\ Lett.\  {\bf 136B} (1984) 387.


\bibitem{Nelson:1984hg}
  A.~E.~Nelson,
  Phys.\ Lett.\  {\bf 143B} (1984) 165.


\bibitem{Barr:1984qx}
  S.~M.~Barr,
  Phys.\ Rev.\ Lett.\  {\bf 53} (1984) 329.


\bibitem{Barr:1984fh}
  S.~M.~Barr,
  Phys.\ Rev.\ D {\bf 30} (1984) 1805.


\bibitem{Bjorkeroth:2015ora}
  F.~Björkeroth, F.~J.~de Anda, I.~de Medeiros Varzielas and S.~F.~King,
  JHEP {\bf 1506} (2015) 141
  [arXiv:1503.03306 [hep-ph]].


\bibitem{Antusch:2013rla}
  S.~Antusch, M.~Holthausen, M.~A.~Schmidt and M.~Spinrath,
  Nucl.\ Phys.\ B {\bf 877} (2013) 752
  [arXiv:1307.0710 [hep-ph]].


\bibitem{Antusch:2013kna}
  S.~Antusch, C.~Gross, V.~Maurer and C.~Sluka,
  Nucl.\ Phys.\ B {\bf 877} (2013) 772
  [arXiv:1305.6612 [hep-ph]].


\bibitem{Ahn:2018cau}
  Y.~H.~Ahn,
  arXiv:1804.06988 [hep-ph].


\bibitem{Froggatt:1978nt}
  C.~D.~Froggatt and H.~B.~Nielsen,
  Nucl.\ Phys.\ B {\bf 147} (1979) 277.


\bibitem{Jaeckel:2013uva}
  J.~Jaeckel,
  Phys.\ Lett.\ B {\bf 732} (2014) 1
  [arXiv:1311.0880 [hep-ph]].


\bibitem{Brivio:2017ije}
  I.~Brivio, M.~B.~Gavela, L.~Merlo, K.~Mimasu, J.~M.~No, R.~del Rey and V.~Sanz,
  Eur.\ Phys.\ J.\ C {\bf 77} (2017) no.8,  572
  [arXiv:1701.05379 [hep-ph]].


\bibitem{Bardeen:1977bd}
  W.~A.~Bardeen and S.-H.~H.~Tye,
  Phys.\ Lett.\  {\bf 74B} (1978) 229.


\bibitem{Srednicki:1985xd}
  M.~Srednicki,
  Nucl.\ Phys.\ B {\bf 260} (1985) 689.


\bibitem{Bardeen:1986yb}
  W.~A.~Bardeen, R.~D.~Peccei and T.~Yanagida,
  Nucl.\ Phys.\ B {\bf 279} (1987) 401.


\bibitem{diCortona:2015ldu}
  G.~Grilli di Cortona, E.~Hardy, J.~Pardo Vega and G.~Villadoro,
  JHEP {\bf 1601} (2016) 034
  [arXiv:1511.02867 [hep-ph]].


\bibitem{Feng:1997tn}
  J.~L.~Feng, T.~Moroi, H.~Murayama and E.~Schnapka,
  Phys.\ Rev.\ D {\bf 57} (1998) 5875
  [hep-ph/9709411].


\bibitem{AlHaydari:2009zr}
  A.~Al-Haydari {\it et al.} [QCDSF Collaboration],
  Eur.\ Phys.\ J.\ A {\bf 43} (2010) 107
  [arXiv:0903.1664 [hep-lat]].


\bibitem{Artamonov:2009sz}
  A.~V.~Artamonov {\it et al.} [BNL-E949 Collaboration],
  Phys.\ Rev.\ D {\bf 79} (2009) 092004
  [arXiv:0903.0030 [hep-ex]].


\bibitem{Adler:2008zza}
  S.~Adler {\it et al.} [E949 and E787 Collaborations],
  Phys.\ Rev.\ D {\bf 77} (2008) 052003
  [arXiv:0709.1000 [hep-ex]].


\bibitem{Buras:2015qea}
  A.~J.~Buras, D.~Buttazzo, J.~Girrbach-Noe and R.~Knegjens,
  JHEP {\bf 1511} (2015) 033
  [arXiv:1503.02693 [hep-ph]].


\bibitem{MarchevskiMoriond2018}
  R.  Marchevski, talk given at Moriond EW, 11 March 2018.
  [URL: https://indico.in2p3.fr/event/16579/contributions/60808/]

\bibitem{Fantechi:2014hqa}
  R.~Fantechi [NA62 Collaboration],
  arXiv:1407.8213 [physics.ins-det].


\bibitem{Gorbahn:2011pd}
  M.~Gorbahn, M.~Patel and S.~Robertson,
  arXiv:1104.0826 [hep-ph].


\bibitem{Brod:2010hi}
  J.~Brod, M.~Gorbahn and E.~Stamou,
  Phys.\ Rev.\ D {\bf 83} (2011) 034030
  [arXiv:1009.0947 [hep-ph]].


\bibitem{Antonelli:2009ws}
  M.~Antonelli {\it et al.},
  Phys.\ Rept.\  {\bf 494} (2010) 197
  [arXiv:0907.5386 [hep-ph]].


\bibitem{Ahn:2009gb}
  J.~K.~Ahn {\it et al.} [E391a Collaboration],
  Phys.\ Rev.\ D {\bf 81} (2010) 072004
  [arXiv:0911.4789 [hep-ex]].


\bibitem{Beckford:2017gsf}
  B.~Beckford [KOTO Collaboration],
  arXiv:1710.01412 [hep-ex].


\bibitem{Ahn:2016kja}
  J.~K.~Ahn {\it et al.} [KOTO Collaboration],
  PTEP {\bf 2017} (2017) no.2,  021C01
  [arXiv:1609.03637 [hep-ex]].


\bibitem{Moulson:2016zsl}
  M.~Moulson [NA62-KLEVER Project Collaboration],
  J.\ Phys.\ Conf.\ Ser.\  {\bf 800} (2017) no.1,  012037
  [arXiv:1611.04864 [hep-ex]].


\bibitem{Buttazzo:2017ixm}
  D.~Buttazzo, A.~Greljo, G.~Isidori and D.~Marzocca,
  JHEP {\bf 1711} (2017) 044
  [arXiv:1706.07808 [hep-ph]].


\bibitem{Ammar:2001gi}
  R.~Ammar {\it et al.} [CLEO Collaboration],
  Phys.\ Rev.\ Lett.\  {\bf 87} (2001) 271801
  [hep-ex/0106038].


\bibitem{Cunliffe:2017cox}
  S.~Cunliffe,
  arXiv:1708.09423 [hep-ex].


\bibitem{Adler:2001xv}
  S.~Adler {\it et al.} [E787 Collaboration],
  Phys.\ Rev.\ Lett.\  {\bf 88} (2002) 041803
  [hep-ex/0111091].


\bibitem{Aubert:2004ws}
  B.~Aubert {\it et al.} [BaBar Collaboration],
  Phys.\ Rev.\ Lett.\  {\bf 94} (2005) 101801
  [hep-ex/0411061].


\bibitem{Grygier:2017tzo}
  J.~Grygier {\it et al.} [Belle Collaboration],
  Phys.\ Rev.\ D {\bf 96} (2017) no.9,  091101
   Addendum: [Phys.\ Rev.\ D {\bf 97} (2018) no.9,  099902]
  [arXiv:1702.03224 [hep-ex]].


\bibitem{delAmoSanchez:2010bk}
  P.~del Amo Sanchez {\it et al.} [BaBar Collaboration],
  Phys.\ Rev.\ D {\bf 82} (2010) 112002
  [arXiv:1009.1529 [hep-ex]].


\bibitem{Abe:2010gxa}
  T.~Abe {\it et al.} [Belle-II Collaboration],
  arXiv:1011.0352 [physics.ins-det].


\bibitem{Patrignani:2016xqp}
  C.~Patrignani {\it et al.} [Particle Data Group],
  Chin.\ Phys.\ C {\bf 40} (2016) no.10,  100001.


\bibitem{Brod:2011ty}
  J.~Brod and M.~Gorbahn,
  Phys.\ Rev.\ Lett.\  {\bf 108} (2012) 121801
  [arXiv:1108.2036 [hep-ph]].


\bibitem{Bai:2018mdv}
  Z.~Bai, N.~H.~Christ and C.~T.~Sachrajda,
  EPJ Web Conf.\  {\bf 175} (2018) 13017.


\bibitem{Li:2018vjr}
  L.~K.~Li,
  Int.\ J.\ Mod.\ Phys.\ Conf.\ Ser.\  {\bf 46} (2018) 1860062.


\bibitem{Rosner:2015wva}
  J.~L.~Rosner, S.~Stone and R.~S.~Van de Water,
  [arXiv:1509.02220 [hep-ph]].


\bibitem{Miskimen:2011zz}
  R.~Miskimen,
  Ann.\ Rev.\ Nucl.\ Part.\ Sci.\  {\bf 61} (2011) 1.


\bibitem{Jaeckel:2017tud}
  J.~Jaeckel, P.~C.~Malta and J.~Redondo,
  arXiv:1702.02964 [hep-ph].


\bibitem{Kitching:1997zj}
  P.~Kitching {\it et al.} [E787 Collaboration],
  Phys.\ Rev.\ Lett.\  {\bf 79} (1997) 4079
  [hep-ex/9708011].


\bibitem{Jodidio:1986mz}
  A.~Jodidio {\it et al.},
  Phys.\ Rev.\ D {\bf 34} (1986) 1967
   Erratum: [Phys.\ Rev.\ D {\bf 37} (1988) 237].


\bibitem{Bayes:2014lxz}
  R.~Bayes {\it et al.} [TWIST Collaboration],
  Phys.\ Rev.\ D {\bf 91} (2015) no.5,  052020
  [arXiv:1409.0638 [hep-ex]].


\bibitem{PerrevoortThesis}
  A.-K. Perrevoort, 
  Sensitivity Studies on New Physics in the Mu3e Experiment and Development of Firmware for the Front-End of the Mu3e Pixel Detector,
  PhD thesis, Heidelberg University, 2018 [unpublished].

\bibitem{Albrecht:1995ht}
  H.~Albrecht {\it et al.} [ARGUS Collaboration],
  Z.\ Phys.\ C {\bf 68} (1995) 25.


\bibitem{Hirsch:2009ee}
  M.~Hirsch, A.~Vicente, J.~Meyer and W.~Porod,
  Phys.\ Rev.\ D {\bf 79} (2009) 055023
   Erratum: [Phys.\ Rev.\ D {\bf 79} (2009) 079901]
  [arXiv:0902.0525 [hep-ph]].


\bibitem{Goldman:1987hy}
  J.~T.~Goldman {\it et al.},
  Phys.\ Rev.\ D {\bf 36} (1987) 1543.


\bibitem{TheMEG:2016wtm}
  A.~M.~Baldini {\it et al.} [MEG Collaboration],
  Eur.\ Phys.\ J.\ C {\bf 76} (2016) no.8,  434
  [arXiv:1605.05081 [hep-ex]].


\bibitem{Baldini:2018nnn}
  A.~M.~Baldini {\it et al.} [MEG II Collaboration],
  Eur.\ Phys.\ J.\ C {\bf 78} (2018) no.5,  380
  [arXiv:1801.04688 [physics.ins-det]].


\bibitem{Bolton:1988af}
  R.~D.~Bolton {\it et al.},
  Phys.\ Rev.\ D {\bf 38} (1988) 2077.


\bibitem{Aubert:2009ag}
  B.~Aubert {\it et al.} [BaBar Collaboration],
  Phys.\ Rev.\ Lett.\  {\bf 104} (2010) 021802
  [arXiv:0908.2381 [hep-ex]].


\bibitem{Adam:2013gfn}
  A.~M.~Baldini {\it et al.} [MEG Collaboration],
  Eur.\ Phys.\ J.\ C {\bf 76} (2016) no.3,  108
  [arXiv:1312.3217 [hep-ex]].


\bibitem{Bellgardt:1987du}
  U.~Bellgardt {\it et al.} [SINDRUM Collaboration],
  Nucl.\ Phys.\ B {\bf 299} (1988) 1.


\bibitem{Blondel:2013ia}
  A.~Blondel {\it et al.},
  arXiv:1301.6113 [physics.ins-det].


\bibitem{DiLuzio:2017ogq}
  L.~Di Luzio, F.~Mescia, E.~Nardi, P.~Panci and R.~Ziegler,
  Phys.\ Rev.\ Lett.\  {\bf 120} (2018) no.26,  261803
  [arXiv:1712.04940 [hep-ph]].


\bibitem{Cirigliano:2017azj}
  V.~Cirigliano, S.~Davidson and Y.~Kuno,
  Phys.\ Lett.\ B {\bf 771} (2017) 242
  [arXiv:1703.02057 [hep-ph]].


\bibitem{Bertl:2006up}
  W.~H.~Bertl {\it et al.} [SINDRUM II Collaboration],
  Eur.\ Phys.\ J.\ C {\bf 47} (2006) 337.


\bibitem{Donghia:2018duf}
  R.~Donghia [Mu2e Collaboration],
  Nuovo Cim.\ C {\bf 40} (2017) no.5,  176.


\bibitem{Wu:2017zwh}
  C.~Wu [COMET Collaboration],
  Nucl.\ Part.\ Phys.\ Proc.\  {\bf 287-288} (2017) 173.


\bibitem{DiBari:2015oca}
  P.~Di Bari and S.~F.~King,
  JCAP {\bf 1510} (2015) no.10,  008
  [arXiv:1507.06431 [hep-ph]].


\bibitem{Antusch:2013jca}
  S.~Antusch and V.~Maurer,
  JHEP {\bf 1311} (2013) 115
  [arXiv:1306.6879 [hep-ph]].


\bibitem{Ade:2015xua}
  P.~A.~R.~Ade {\it et al.} [Planck Collaboration],
  Astron.\ Astrophys.\  {\bf 594} (2016) A13
  [arXiv:1502.01589 [astro-ph.CO]].
\end{thebibliography}
\end{document}